\documentclass[reprint,nofootinbib, amsmath,amssymb,aps,twocolumn]{revtex4-2}
\usepackage{amsmath} % <-- important line, maybe need to put this before other usepackage so that i can use \( \) for inline and \[ \] for equations. 
\usepackage{amssymb,amsmath,amsfonts,amsbsy,graphicx}
\usepackage{mathtools}% http://ctan.org/pkg/mathtools

\usepackage{comment}
\usepackage{epstopdf}
\usepackage{color}
\usepackage{psfrag}
\usepackage{stackrel}

\usepackage{csquotes}
\usepackage{tikz}
\usetikzlibrary{positioning}
\usetikzlibrary{intersections}
\usepackage{siunitx}
\newcommand\be{\begin{equation}}
\newcommand\ee{\end{equation}}
\newcommand\ba{\begin{eqnarray}}
\newcommand\ea{\end{eqnarray}}\newcommand\eq{\begin{equation}}           
\newcommand\en{\end{equation}}

  \newcommand\bcr{\begin{comment}}  % begin comment-region
  \newcommand\ecr{\end{comment}} % end comment-region

 \newcount\colveccount
\newcommand*\colvec[1]{
        \global\colveccount#1
        \begin{pmatrix}
        \colvecnext
}
\def\colvecnext#1{
        #1
        \global\advance\colveccount-1
        \ifnum\colveccount>0
                \\
                \expandafter\colvecnext
        \else
                \end{pmatrix}
        \fi
}
\input epsf

\usepackage{ulem}
\usepackage{amssymb}
\usepackage{amsmath}
\usepackage{bm}
\usepackage{graphicx}
\usepackage{color}

\usepackage{caption}
\usepackage{subcaption}
\def\gsim{\;\rlap{\lower 2.5pt
 \hbox{$\sim$}}\raise 1.5pt\hbox{$>$}\;}
\def\lsim{\;\rlap{\lower 2.5pt
 \hbox{$\sim$}}\raise 1.5pt\hbox{$<$}\;}

\usepackage{graphicx}

\usepackage{url}  % Make sure to include this package to put web address in bibliography
\usepackage{amssymb,amsmath}
\usepackage{amsbsy} 

\newcommand{\apjl}{ApJL}

\newcommand{\mnras}{MNRAS}

  \newcommand\diracode{.}
 
\usepackage{graphicx}
\usepackage{url}
\usepackage{amssymb,amsmath}
\usepackage{amsbsy}

\usepackage{varioref} % Equations \vrefrange{eq1}{eq5}
\usepackage[compress]{cleveref}

\begin{document}
\title{
  %{\hfill  \small\\ ~\\~\\
  SKA Sensitivity to Sub-GeV Dark Matter Decay: Synchrotron Radio Emissions in White Dwarf Magnetospheres
}
\author{Kenji Kadota$^{1,2}$ and Shota Kisaka$^{3}$\\
  {\small  $^1$School of Fundamental Physics and Mathematical Sciences,
  Hangzhou Institute for Advanced Study,\\ University of Chinese Academy of Sciences (HIAS-UCAS), Hangzhou 310024, China\\
  $^2$International Centre for Theoretical Physics Asia-Pacific (ICTP-AP), Beijing/Hangzhou, China
  \\
  $^3$Physics Program, Graduate School of Advanced Science and Engineering,\\Hiroshima University, Higashi-Hiroshima, 739-8526, Japan
  }
}
%\date{\vspace{-5ex}}
% \date{}  % this simply makes the date blank. Toggle commenting to test

\begin{abstract}
  % We study synchrotron radiation in the magnetosphere of a white dwarf, considering the relativistic electrons/positrons from the decay of dark matter forming an ultra-compact minihalo around a primordial black hole.
We investigate the potential of the Square Kilometre Array (SKA) in detecting synchrotron radiation emitted from the decay of sub-GeV dark matter (dark matter with masses below the GeV scale) in the presence of strong magnetic fields. As a concrete setup, we consider scenarios where the magnetosphere of a magnetic white dwarf overlaps with dense dark matter environments, such as those surrounding a primordial black hole. Our study reveals that the encounters of compact objects such as white dwarfs and black holes offer a promising avenue for upcoming radio telescopes to probe the properties of light dark matter, which has been less explored compared with more conventional heavier (masses above the GeV scale) dark matter.

\end{abstract}

\maketitle   

\setcounter{footnote}{0} 
\setcounter{page}{1}\setcounter{section}{0} \setcounter{subsection}{0}
\setcounter{subsubsection}{0}

\section{Introduction}
The nature of the dark matter, such as its mass and interaction, remain elusive despite the compelling supports for its existence from the astrophysical observations. While many conventional dark matter search experiments have targeted weak-scale dark matter (typical mass scale is of order $\sim 100$ GeV), the lack of observational signals has expanded our interest in exploring a broader range of potential dark matter candidates. The possible dark matter mass can indeed span a wide spectrum, from ultra-light axion-like particles with masses even below ${\cal O}(10^{-20})$ eV to scales surpassing the Planck scale, such as primordial black hole dark matter \cite{Jungman:1995df,Arcadi:2017kky,Boddy:2022knd,Chou:2022luk,Carr:2020xqk,Chadha-Day:2021szb}.

Our paper focuses on the decay of sub-GeV dark matter (masses below 1 GeV scale), which has not been fully explored in comparison to heavier dark matter \cite{Knapen:2017xzo,Battaglieri:2017aum,Bondarenko:2019vrb,Coogan:2019qpu,Essig:2013goa,Cirelli:2020bpc,Choudhury:2019tss,Hooper:2007tu,Tashiro:2014tsa,Xu:2018efh,Ooba:2019erm,Boehm:2002yz,Cirelli:2023tnx}. Specifically, we are investigating synchrotron radiation signals in dense dark matter environments in the presence of strong magnetic fields. More concretely, we consider the ultra-compact minihalo around a primordial black hole as a dense dark matter environment, and the magnetic white dwarf as the source of a strong magnetic field.

The dark matter can accrete into a black hole to form the ultra compact minihalo and their consequent decay or annihilation from such a large density region has been studied for the detection, for instance, by the gamma-ray, CMB (Cosmic Microwave Background), radio and neutrino telescopes \cite{Carr:2023tpt,Freese:2022ouh,Kuhnel:2018kwf,Scott:2009tu,Gondolo:1999ef,Lacki:2010zf,Adamek:2019gns,Boucenna:2017ghj,Eroshenko:2016yve,Carr:2020mqm,Cai:2020fnq,Delos:2018ueo,Kohri:2014lza,Bertone:2019vsk,Ando:2015qda,Hertzberg:2020kpm,Yang:2020zcu,Zhang:2010cj,Tashiro:2021xnj,Yang:2020zcu,Kadota:2022cij}. The effects of the dense dark matter environment on gravitational wave signals also have been studied, for instance, through the de-phasing effects of the gravitational waveform by the LISA-like satellite experiments \cite{Eda:2013gg,Eda:2014kra,Kadota:2020ahr, Yue:2017iwc,Macedo:2013qea,Barausse:2014tra,Bertone:2019irm,Cole:2022fir,Kavanagh:2020cfn,Cardoso:2019rou,Hannuksela:2019vip,Coogan:2021uqv,Kim:2022mdj,Kadota:2023wlm}.

%% We study in this paper the synchrotron radiation signals from the dense dark matter environments in the presence of the strong magnetic fields. In particular we focus on the sub-GeV (the mass below 1 GeV) dark matter.
%% Many previous relevant literature discussing the indirect dark matter search studied the dark matter with the mass above GeV scale (a typical example is the weakly interacting massive particles (WIMPs) with the weak scale mass (of order $\sim 10^{2-3}$ GeV). The heavier dark matter produces more energetic final products from the decay and annihilation and they exit tight bounds on heavy (the mass above GeV) dark matter for instance from Fermi-LAT gamma-ray telescope \cite{}.
%% Many dark matter direct search experiments which measure the recoil energy from the nuclei target also focus on the heavier dark matter and lose the sensitivity for the dark matter mass below 1 GeV scale \cite{}. Facing the lack of any observational hints of conventional weak scale mass dark matter, such as WIMPs, growing attention has been paid to light (sub-GeV) DM \cite{Kuhnel:2018kwf, Knapen:2017xzo,Battaglieri:2017aum,Bondarenko:2019vrb,Coogan:2019qpu,Essig:2013goa,Cirelli:2020bpc,Choudhury:2019tss,Hooper:2007tu,Tashiro:2014tsa,Xu:2018efh,Ooba:2019erm}.

One of reasons why the light (sub-GeV) dark matter is preferable for our signals is that heavier dark matter leads to the peak frequency of synchrotron radiation signals (the peak frequency is proportional to the magnetic field times square of charged particles' Lorentz factor $\propto B \gamma^2$) too big for the SKA \cite{skawebpage,2019arXiv191212699B} sensitive frequency range (around $50 MHz-50 GHz$). Another notable difference from the previous works is that we consider the strong magnetic fields from the magnetic white dwarfs. Many previous works studying the synchrotron radiation signals from the dark matter decay and annihilation considered the heavy (mass above GeV scale) dark matter in the presence of the magnetic fields typically of order ${\cal O} (1-10)\mu$G including those from Galactic center, globular clusters, dwarf galaxies and galaxy clusters \cite{Cembranos:2019noa,Colafrancesco:2015ola,Chen:2021rea,Wang:2023sxr,Ghosh:2020ipv,Colafrancesco:2005ji,McDaniel:2017ppt}. For instance Ref. \cite{Kadota:2022cij} studied synchrotron radiation from the ultra compact minihalos surrounding black holes in the presence of the Galactic magnetic fields. The synchrotron radiation is indeed the dominant radiation for the weak scale dark matter for the magnetic fields of order ${\cal O} (1-10)\mu$G, which leads to the tight bounds on the abundance of ultra compact minihalos from the currently available radio observation data. For the sub-GeV dark matter mass with such small magnetic fields, due to the smaller energy of final products emitted by the dark matter, the synchrotron radiation power amplitude and peak frequency become too low to be observable by the radio telescopes. See, for instance, Ref. \cite{Dutta:2020lqc} for the scenarios where the inverse Compton scattering, which has a higher peak frequency than that of synchrotron radiation, can instead lead to the radio frequency flux detectable by the SKA for sub-GeV dark matter with order ${\cal O} (1-10)\mu G$ magnetic fields.

In our paper, we focus on the significantly larger magnetic fields of the order ${\cal O}(10^{3-7})$G from magnetic white dwarfs. With such large magnetic fields, sub-GeV dark matter can lead to the radio frequency signals within a reach of upcoming SKA radio telescope. In the presence of these intense magnetic fields, the other background photon energy becomes negligible compared with the magnetic field energy, and the synchrotron emission can be the dominant radiation mechanism at radio frequencies. For even higher magnetic fields such as those from the neutron star magnetosphere, however, the peak frequency of the synchrotron radiation can be too high for the SKA sensitive frequency range in our scenarios. Additionally, the larger volume of the magnetosphere of a white dwarf, from which the signals originate, compared to that of a neutron star, is also advantageous for our scenarios.

Our paper is structured as follows. Section \ref{sec:setup} outlines our formalism to estimate the synchrotron radiation from the dark matter decay, and illustrates the realization of dense dark matter environment surrounding a primordial black hole. Section \ref{sec:results} presents our findings on the lower bounds of the dark matter decay rate required for detectability by the forthcoming SKA radio telescope.
\section{Setup}
\label{sec:setup}
We consider the synchrotron radiation from the relativistic electrons in the presence of  strong magnetic fields.
As a source of the relativistic electrons, we consider the decaying dark matter.
For the strong magnetic fields, we consider the magnetic white dwarfs.
In addition, our scenarios assume the dense dark matter environments. We hence also briefly discuss the ultra compact minihalo surrounding a primordial black hole as a concrete realization for large dark matter density. 
\subsection{Synchrotron radiation from dark matter decay}
The synchrotron radiation flux density received on the Earth is
% One Jansky is equal to 
%$10^{-26}$ watts per square meter per hertz.
\ba
S_{\nu} = \frac{1}{D^2} \int_{R_1}^{R_2}  dr r^2 j_{\nu}
\label{eq:snujy}
\ea
$D$ is the luminosity distance from the source of the synchrotron radiation to the Earth.
The emissivity of the synchrotron radiation source (the energy emitted per unit volume, per unit frequency, per unit time) is
\ba
j_{\nu}(\nu,r) %[cm^{-3} W Hz^{-1}]
=2 \int^{m_{\chi}/2}_{m_e c^2}
dE
\frac{dn_e}{dE}(E) P_{\nu}(\nu,E,r)
\ea
%The integration of $j_{\nu}$ calculates the total power emitted per unit frequency per unit volume over a spherical volume of the source.
We for concreteness consider the magnetic white dwarf and the limits of integration, $R_1,R_2$, represent the inner and outer radii of the white dwarf magnetosphere. A factor $2$ accounts for the contribution both from electrons and positrons. The power radiated by an electron per unit frequency for synchrotron radiation is given by
\ba
P_{\nu} = \frac{\sqrt3  e^3  B(r)  \sin(\alpha)}{ m_e  c^2 } F(x)
\ea
where $\alpha$ is the pitch angle (the angle between the velocity vector of the electron and the magnetic field).
$F(x)$ is the synchrotron function (the low frequency part is well approximated by the power law of 1/3 and decays exponentially well above the critical frequency)
\ba
F(x) = x  \int_x^{\infty} K_{5/3}(\xi) d\xi
\ea
$K_{5/3}$ is the modified Bessel function of the second kind of order 5/3, and $x = \nu / \nu_c$, where $\nu_c= 3  e  B  \gamma^2  \sin(\alpha) / (4  \pi  m_e  c)$ is the critical frequency.
%% =4.2\times 10^{25} \left( \frac{B}{10^{10}G} \right)
%% \left( \frac{ \gamma }{10^5} \right)^2
%% \left( \frac{ \sin (\alpha) }{0.1} \right)
%% [Hz]
%% \ea
We assume the simple dipole profile $B(r)\propto r^{-3}$ for the white dwarf magnetic field. The electron energy spectrum $\frac{dn_e}{dE}$ (the electron number density per unit energy) arising from the dark matter decay can be estimated as
\ba
\frac{dn_e}{dE}
\approx
%\Gamma^{total}_{\chi}\times min[t_{loss},t_{ad}] \times   dN_e /dE
n_{\chi} \Gamma_{\chi}  \times min[t_{cool},t_{ad}]  \times  \frac{dN_e}{dE}  
\ea
%% where total decay rate per volume is defined as the dark matter number density $n_{\chi}$ times the dark matter decay rate $\Gamma_{\chi}$ %\cite{Kuhnel:2018kwf}:
%% \ba
%% \Gamma^{total}_{\chi}=n_{\chi} \Gamma_{\chi}
%% \ea
where $n_{\chi},\Gamma_{\chi}$ are respectively the dark matter number density and decay rate.
$dN_e/dE$ represents the electron energy distribution from each dark matter decay for a given final state channel. We assume for illustration the dark matter decays dominantly into electron positron pair final states $\chi\rightarrow e^+ e^-$.
Two timescales $t_{cool}, t_{ad}$ are the cooling timescale and the advection timescale.
The cooling timescale of synchrotron radiation can be estimated by
\ba
t_{cool}
&\sim&\frac{\gamma m_e c^2 \sin \alpha}{P_{syn}}
%{\frac{2e^4 B(r)^2 \gamma^2 \beta^2 \sin ^2 \alpha}{3m_e^2 c^3}}
%\frac{E}{P^{total}_{synchrotron}(r)}
%% \sim
%% \\
%% \frac{\gamma m_e c^2}{\frac{2e^4 B(r)^2 \gamma^2 \beta^2 \sin ^2 \alpha}{3m_e^2 c^3}}
\\
&\sim& 4 \times 10^{-5}
\left(
\frac{10^5 G}{B(r)}
\right)^2
\left(
\frac{10^2}{\gamma}
\right)
\left(
\frac{1}{\sin \alpha}
\right) [s]
\ea
$\gamma$ is the Lorentz factor of a radiating charged particle and $P_{syn}$ is the total power (integrated over all frequency) of the synchrotron radiation for each particle
%We can compare this with the relativistic case $P_{syn}$ integrated over all frequency (most of the signal comes from $\nu=\nu_{c}$?)
\ba
P_{syn}=
\frac{2e^4 B(r)^2 \gamma^2 \beta^2 \sin ^2 \alpha}{3m_e^2 c^3}
%% =
%% 2.0 \times 10^{-4} erg/s
%% \left( \frac{B}{10^5 G} \right)^2
%% \left( \frac{\gamma}{10} \right)^2
%% \left( \frac{\beta}{1} \right)^2
%% \left( \frac{\sin \alpha}{0.1} \right)^2
\ea
The advection timescale at the emission region $r$ can be estimated by
\ba
t_{ad}\sim \frac{r}{c} \sim 0.2 \left( \frac{r}{0.1 R_{\odot}}\right) [s]
\ea
When we integrate over the white dwarf magnetosphere, the magnetic field is bigger for a smaller radius while there is a bigger volume contribution for a bigger radius. The relevant timescale, the minimum of $t_{cool}$ and $t_{ad}$, becomes biggest at a radius where $t_{cool} \approx t_{ad}$.

%% We consider the timescale min(tadvection, tloss) also look at kisaka paper \cite{Kisaka:2017rpz}
%% At the emission region $r$

%% We can see that, for the large magnetic field, the synchrotron radiation energy loss is instantaneous compared with the advection timescale.

%% We assume all DM particles decay dominantly to $e^+e^-$.
%% I assumed the radiation energy loss is instantaneous at the position $r$ and dominated by the synchrotron radiation.

%\subsection{Magnetic white dwarfs}
We consider the magnetic white dwarf as the source of the strong magnetic field.
The magnetic white dwarfs (MWDs) are characterized by the strong magnetic fields and they can make up of order ${\cal O}(10)$\% of the white dwarf population \cite{Ferrario:2015oda,2020AdSpR..66.1025F,2022ApJ...935L..12B,Ferrario:2015oda}. The magnetic fields of MWDs can typically range from $10^3$G to $10^9$G, and some of closest known MWDs are within 20 parsecs of the Earth.
In our quantitative discussions in Section \ref{sec:results}, we choose the distance to the white dwarf $D=100pc$, its radius $R_1=0.01 R_{\odot}$, the size of the magnetosphere $R_2=10\times R_1$, the pitch angle $\sin \alpha=1$ as the fiducial values for illustration. We will demonstrate, in Section \ref{sec:results}, that a magnetic field of around $10^6$G on the surface of a white dwarf represents an optimal value for our signal estimates. Larger magnetic fields result in signal peak frequencies falling outside the SKA sensitivity window, despite an increase in signal amplitude with stronger magnetic fields \footnote{We also mention that the magnetosphere of a neutron star could be an alternative possibility for our setups. We however found that the magnetosphere of a white dwarf is more suitable to our scenarios than that of a neutron star, due to its relatively weaker magnetic field (ensuring signal frequencies within the SKA's sensitivity range) and its larger volume (allowing for greater spatial overlap with dark matter).}.

\subsection{Dense dark matter environment around a primordial black hole}
The signals of indirect dark matter search can be enhanced in the presence of the dense dark matter environment. As a concrete example to realize such a high density dark matter region, we illustrate the ultra-compact minihalo surrounding a primordial black hole. We adopt the analytical expression of Refs.\cite{Carr:2020mqm,Boudaud:2021irr,Eroshenko:2016yve,Boucenna:2017ghj,Chanda:2022hls} which is given by (for brevity, we use the convention $c=1$ in the following discussions unless stated otherwise) \footnote{
For concreteness, we assume a primordial black hole rather than an astrophysically sourced black hole. The purpose of this section is simply to illustrate an example of a large dark matter density. The discussions on radio emissions in subsequent sections, while focused on this context, can also qualitatively apply to other environments with dense dark matter.}
%For concreteness, we assume the primordial black hole in our illustration rather than an astrophysically sourced (stellar collapse) black hole for its simplicity. Our goal in this section is merely to illustrate the existence of the large dark matter density and our radio emission discussions in the following sections can be qualitatively applicable to other dense dark matter environments as well.}
\begin{equation}
\rho_{\chi}(x) =
\begin{cases}
 \rho_{kd} \left( \frac{r_c}{r} \right)^{3/4} & \text{for } r \leq r_c \\
 \frac{\rho_{eq}}{2} \left( \frac{M}{M_{\odot}} \right)^{3/2} \left( \frac{\hat{r}}{r} \right)^{3/2} & \text{for } r_c < r \leq r_k \\
  \frac{\rho_{eq}}{2} \left( \frac{M}{M_{\odot}} \right)^{3/4} \left( \frac{\bar{r}}{r} \right)^{9/4} & \text{for } r > r_k
\end{cases}
\end{equation}
where
\ba
\hat{r}\equiv GM_{\odot}\frac{t_{eq}}{t_{kd}} \frac{m_{\chi}}{T_{kd}}, 
\bar{r}\equiv (2GM_{\odot}t^2_{eq})^{1/3}
\ea
The transition radii are given by
\ba
r_c \sim \frac{r_{sch}}{2} \left( \frac{m_{\chi}}{T_{kd}}\right),
r_K \sim 4 \frac{t^2_{kd}}{r_{sch}} \left( \frac{T_{kd}}{m_{\chi}} \right)^2
\ea
where $r_{sch}=2GM$ is the Schwarzschild radius
%$2GM=2.9(M/M_{\odot})[km]$
for a given black hole mass $M$. $\rho_{eq},t_{eq}$ are respectively the density of the Universe and the time at the matter-radiation equality epoch. $\rho_{kd},t_{kd}$ are those at the dark matter kinetic decoupling. The epoch when the dark matter kinetically decouples from the cosmic plasma affects the velocity of dark matter, and it can influence the dark matter profile because the dark matter kinetic energy can play an important role in how it is trapped by a black hole's gravitational field. The dark matter kinetic decoupling temperature $T_{kd}$ is given by \footnote{The exact nature of the kinetic decoupling can be heavily model-dependent. We refer the readers to, for instance, Refs. \cite{Loeb:2005pm,Bertschinger:2006nq,Gondolo:2012vh, Profumo:2006bv,Gondolo:2016mrz,Green:2003un,Green:2005fa,Bringmann:2006mu} for more detailed physics and the phenomenology of dark matter kinetic decoupling.}
%The conventional kinetic decoupling temperature $T_{KD}\propto m^{5/4}$ we adopted for concreteness is a typical scaling for the bino-like neutralino dark matter, and we refer the readers to for instance Refs. \cite{Loeb:2005pm,Bertschinger:2006nq,Gondolo:2012vh, Profumo:2006bv,Gondolo:2016mrz,Green:2003un,Green:2005fa,Bringmann:2006mu} for more detailed physics and the phenomenology of dark matter kinetic decoupling.}
\ba
T_{kd}=\frac{m_{\chi}^{5/4}}{\Gamma[3/4]}\left( \frac{\alpha }{M_{pl}}\right)^{1/4}
\ea
and the corresponding kinetic decoupling time is obtained from the Friedmann equation %($H=(2t)^{-1}=\alpha T^2 m_{pl}^{-1}$).
\ba
t_{kd}=\frac{m_{pl}}{2 \alpha T_{kd}^2}%\sim 2.40391\times 10^{21}/GeV
\ea
with $\alpha=(16 \pi^3 g_{kd}/45)^{1/2}$. For concreteness, we choose the relativistic degrees of freedom at the kinetic decoupling $g_{kd}=61.75$ \cite{Bringmann:2006mu,Boucenna:2017ghj,Adamek:2019gns}. $\Gamma$ is the gamma function. % $\Gamma(3/4)$.%=1.2254167024651779.
%$T_{KD}\propto m^{5/4}$ is a typical scaling for the bino-like neutralino dark matter, and,
  We refer the readers to Refs.\cite{Carr:2020mqm,Boudaud:2021irr,Eroshenko:2016yve,Boucenna:2017ghj,Chanda:2022hls,Vasiliev:2007vh,Sadeghian:2013laa,Eroshenko:2016yve} for more detailed discussions on the dark matter profile.

\begin{figure}[!htbp]
     \begin{tabular}{c}
              \includegraphics[width=0.5\textwidth]{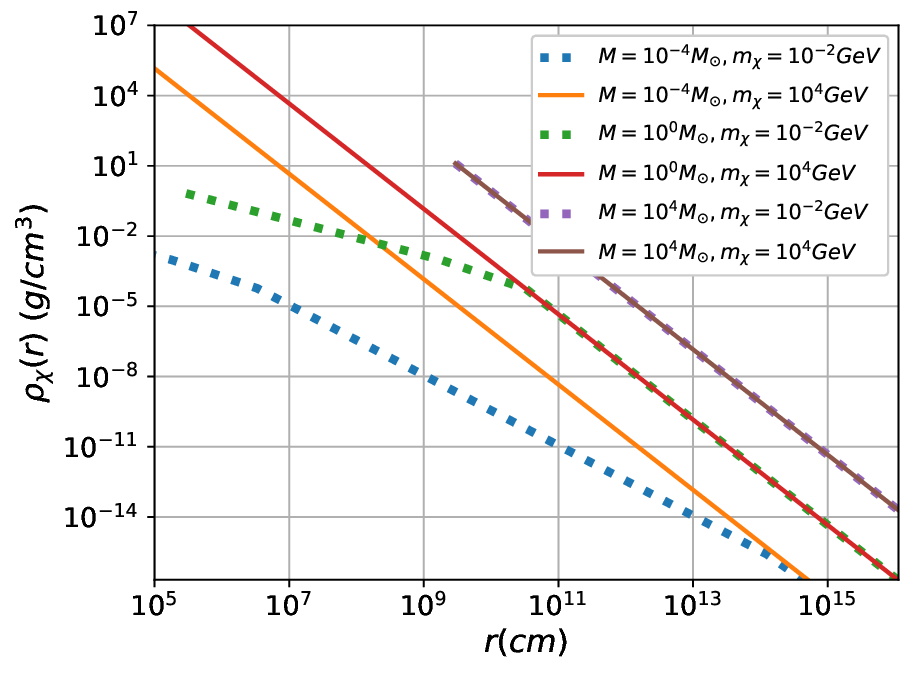}
       % &
 %      \includegraphics[width=0.5\textwidth]{
 %        \diracode/dmrhor2} 
           \end{tabular}
     % inkscape input.pdf --export-eps=output.eps  (uses inkscape) <- works best
     % pdftops -eps my.pdf my.eps (uses poppler)
  %\vspace{-25mm}
     \caption{The enhanced dark matter density around a primordial black hole for different black hole masses $(M=10^{-4},1,10^4M_{\odot})$ and dark matter masses $(m_{\chi}=10^{-2}GeV, 10^4 GeV)$. The curve is truncated at Schwarzschild radius. Two curves for $M=10^4M_{\odot}$ with $m_{\chi}=10^{-2}$ and $m_{\chi}=10^4$ GeV are indistinguishable.}
     %%  Made by /Users/kenji/BGD/work/codes/kisakaPulsarRadiation/oct17rhor.ipynb
 %  pdflatex -shell-escape draft1.tex  ->  bibtex draft1 -> pdflatex -shell-escape draft1
   \label{fig:dmrhor}
\end{figure}

The profiles are illustrated in Fig. \ref{fig:dmrhor}. The curve is truncated at the Schwarzschild radius. % \footnote{The relativistic effects on the dark matter orbit can become important for a small radius ($\lesssim 2 r_{sch}$), and the dark matter orbit can be unstable close to the Schwarzschild radius ($\lesssim 3 r_{sch}$) \cite{Vasiliev:2007vh,Sadeghian:2013laa,Eroshenko:2016yve}. The region of our interest would be hence in the range at least bigger than these scales.}. The conventional `spike' profile $\rho(r)\propto r^{-9/4}$ shows up for the dark matter halo surrounding a primordial black hole. %%%%%%%%%%%%%%%
%%%%%%%%%%%%%%%%
%%%%%%%%%%%%%%%%%%%%%%%%
%%%%%%%%%%%%%%%%%%%%
Such a profile can be obtained analytically assuming the adiabatic growth of a halo and it has also been verified numerically \cite{Gondolo:1999ef,Adamek:2019gns,Serpico:2020ehh} \footnote{The power for the spike profile $\rho(r) \propto r^{-\gamma_{sp}}$ for a given initial profile $\rho_{ini}\propto r^{-\gamma_{ini}}$ can be analytically derived as $\gamma_{sp}=(9-2\gamma_{ini})/(4-\gamma_{ini})$. For instance, $\gamma_{sp}=9/4$ for the spike profile around a primordial black hole with the initial background $\gamma_{ini}=0$ deep in the radiation dominated epoch, and $\gamma_{sp}=7/3$ if the halo growth starts from the cuspy NFW profile $\gamma_{ini}=1$. A steeper slope can lead to a bigger density and some literature discusses even a value up to $\gamma_{sp}\sim 2.75$ \cite{Gondolo:1999ef,Ullio:2001fb,Shapiro:2022prq,Fields:2014pia,Shapiro:2016ypb}. Our qualitative discussions and conclusions are not affected by those choices of parameters and we show the density profile for $\gamma_{sp}=9/4$ in Fig. \ref{fig:dmrhor} to illustrate a possible realization of a large dark matter density.}.
For the radius below $r_k$ (which characterizes the scale inside which the dark matter particle kinetic energy is bigger than its potential energy under the influence of the gravitational filed of the central object (such as a black hole)), the slope becomes milder. The halo profile for such an inner region can be obtained analytically by requiring the phase space conservation assuming Maxwell-Boltzmann distribution for the dark matter velocity \cite{Carr:2020mqm,Boudaud:2021irr,Eroshenko:2016yve,Boucenna:2017ghj,Chanda:2022hls}.%, and the ultra compact minihalo halo profile is illustrated in Fig. \ref{fig:dmrhor}.

%The bigger black hole mass has a bigger Schwarzschild radius and the minimum radius shown in the plot corresponds to the Schwarzschild radius $r_S=2GM$ for a given BH mass $M$

%For some parameter range of black hole mass and dark matter mass, the big dark matter halo density can be realized for a smaller black hole mass by approaching to a smaller radius.

Having illustrated how big the dark matter density can be around a primordial black hole, which even could reach the values be as big as ${\cal O}(1) g/cm^3$, we simply adopt the value $\rho_{\chi}=1g/cm^3$ as the fiducial value for quantitative discussions in Section \ref{sec:results}. Readers can straightforwardly re-scale our calculation results for a different value because our synchrotron radiation signal (Eq.~(\ref{eq:snujy})) scales linearly with $\rho_{\chi}$. We mention that the dynamics of white dwarf motion in a high dark matter density environment, especially near a black hole, can be complex. Our objective is to explore the potential effects of such an environment, serving as a proof of concept for the parameters we have adopted. Our preliminary investigation aims to understand the implications of a white dwarf encountering a region rich in dark matter. In our present analysis, we disregard variations in dark matter density, since the synchrotron radiation is promptly emitted in strong magnetic fields. This suffices for our goal of demonstrating the potential significance of our scenarios for the future radio observations.

\section{Results}
\label{sec:results}
We compare the possible signals in our scenarios with the forthcoming SKA sensitivity \cite{skawebpage,2019arXiv191212699B}. %,Cembranos:2019noa,2019arXiv191212699B,Chen:2023fgr}.
The SKA sensitivity is estimated by the radiometer equation
\ba
  S_{min}  =
  \frac{2k_bT_{sys}}{A_{eff} \eta_s \sqrt{\eta_{pol} t_{obs} \Delta B}}
%    \frac{2k_bT_{sys}}{A_{eff} \eta_s \sqrt{ t_{obs} \Delta B}}
  \ea
  %% \ba
  %%   S_{min}  \sim 
  %% {\bf check the numbers}
  %% \sim  0.205152 Jy\\
  %% \left( \frac{A_{eff}/T_{sys}}{10^3 m^2/K}\right)^{-1}
  %%   \left(\frac{t_{obs}}{100 hrs}\right)^{-1/2}
  %%   \left(\frac{\nu}{1 GHz}\right)^{-1/2}
  %%  \ea
  %%  ($k_b=c=\hbar=1$, I substitute etapol=2, delta B=0.3 nu, etas=0.9.
  %%  Jy=2.473e-48 GeV**3)
$k_b$ is the Boltzmann constant, $\eta_{pol}$ is the number of polarization states, $t_{obs}$ is the integrated observation time. $A_{eff}$ is the effective collecting area of the telescope and $T_{sys}$ is the system temperature consisting of the sum of sky/instrumental noises of the system. Note their ratio $A_{eff}/T_{sys}$, so-called the natural sensitivity, is frequency dependent and the values are adopted from Ref. \cite{2019arXiv191212699B}. $\eta_s$ is the system efficiency and $\Delta B$ is the bandwidth. We use $\eta_s=0.9, \eta_{pol}=2$ and adopt the band width which is frequency dependent as 0.3 times the frequency $\Delta B=0.3 \nu$ \cite{2019arXiv191212699B}.
The anticipated sensitivity of the upcoming SKA is expected to cover the frequency range $50MHz\sim 50 GHz$ (combining SKA-Low (covering the lower frequency bands) and SKA-Mid (covering the mid-range frequencies)). 
%The SKA sensitivities for $t_{obs}=100$ and $1000$ hours are illustrated in Fig. \ref{fig1:nov17espectruma}.

%%%%%%%%%%%%%%%%%%%%%%%%
%%%%%%%%%%%%%%%%%%%%%%%%%

\begin{figure}[ht]
    \centering
    \begin{subfigure}[b]{0.48\textwidth}
         \includegraphics[width=\textwidth]{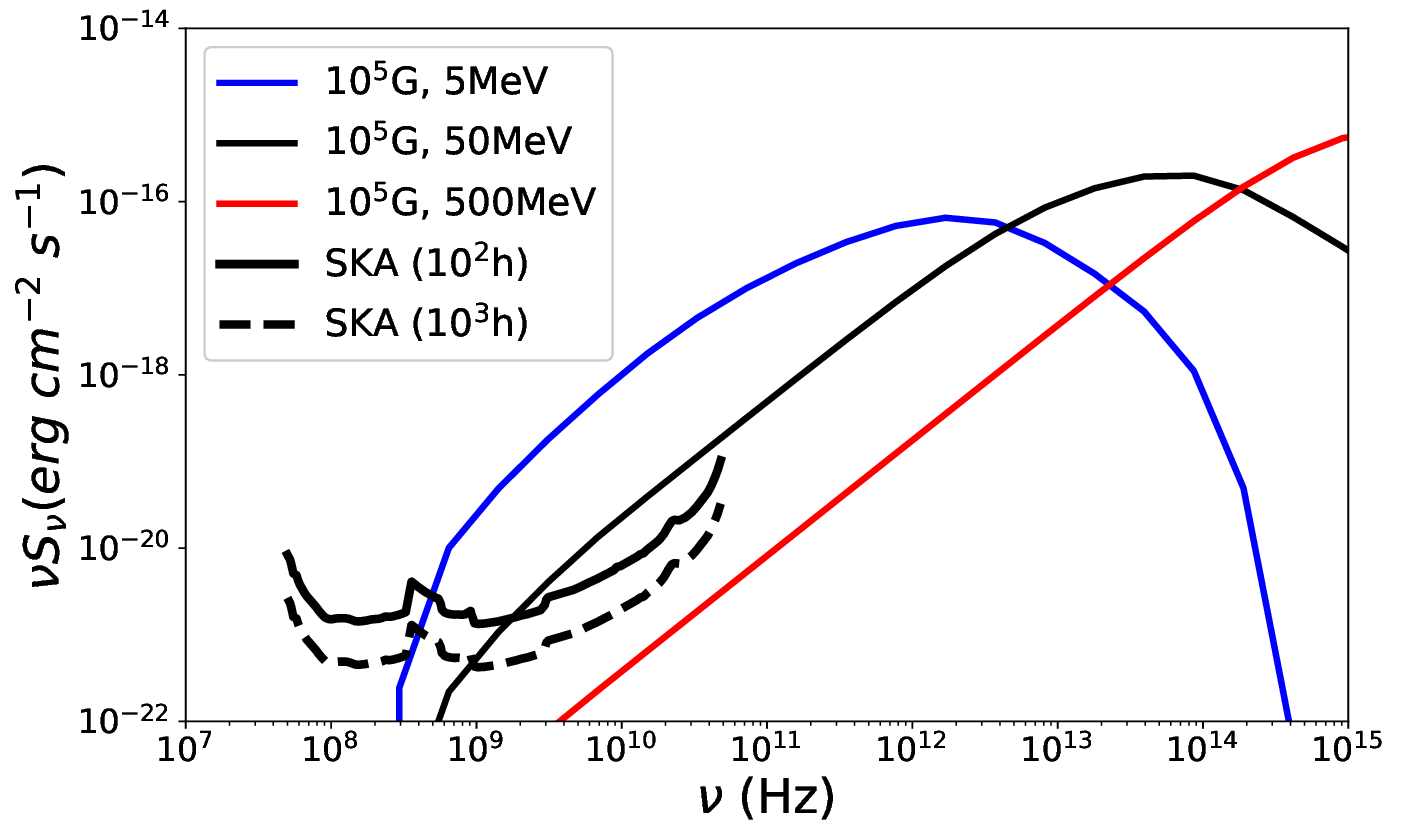}
    \end{subfigure}
    \hfill
    \begin{subfigure}[b]{0.48\textwidth}
                 \includegraphics[width=\textwidth]{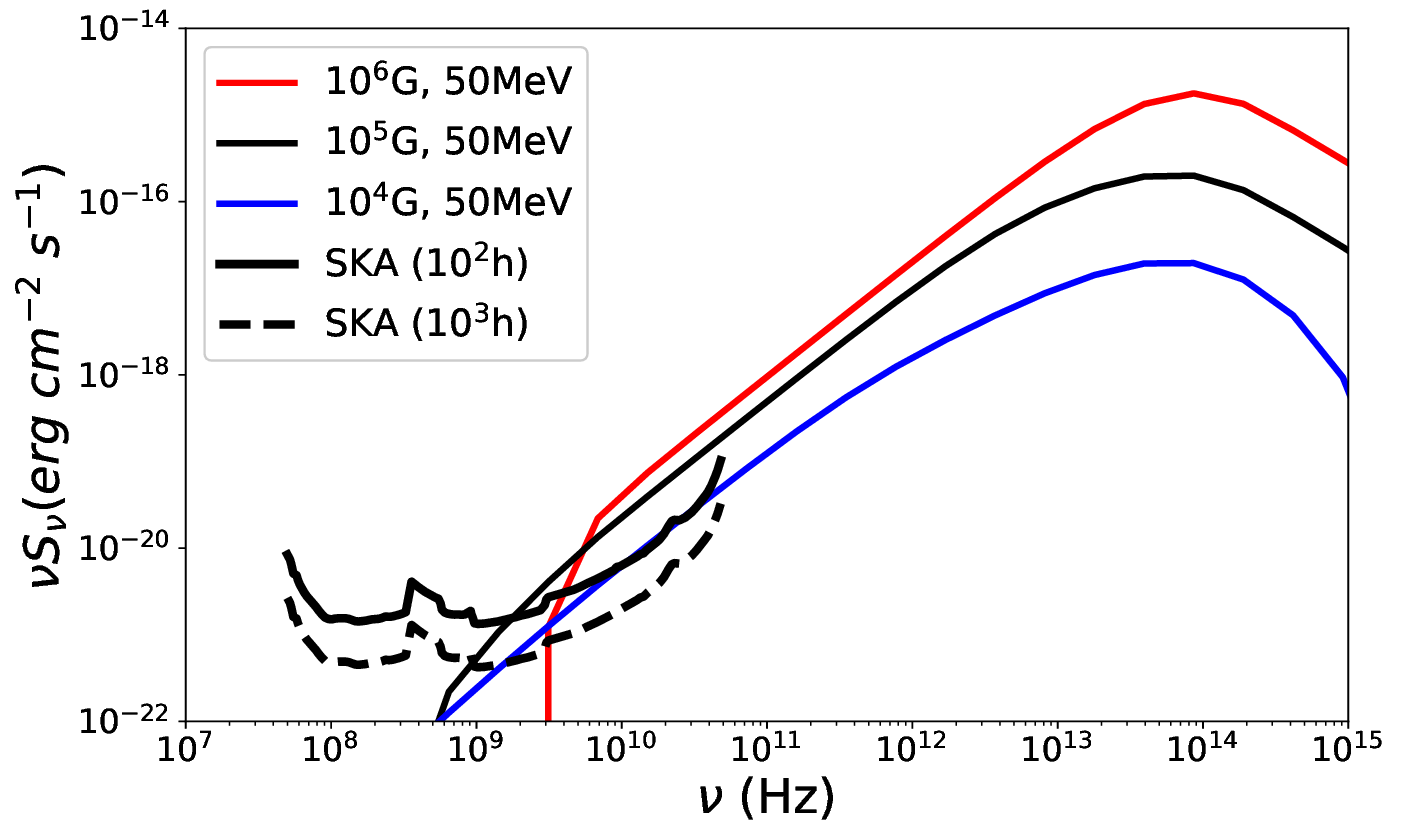}
    \end{subfigure}

       % inkscape input.pdf --export-eps=output.eps  (uses inkscape) <- works best
     % pdftops -eps my.pdf my.eps (uses poppler)
  %\vspace{-25mm}
    \caption{
      The energy spectrum distribution $\nu S_{\nu}$ for the synchrotron radiation as a function of the frequency.
 The SKA sensitivity assuming 100 and 1000 hours of observations are shown for reference.
 The magnetic white dwarf is assumed to be 100 pc away from the Earth. We assume that the magnetic field follows the dipole profile $B(r)=B_0 (r/r_0)^{-3}$ and the magnetosphere spans from $r_0=0.01 R_{\odot}$ to $10 r_0$. The dark matter decay rate $\Gamma_{\chi}=10^{-25}[s^{-1}]$ and the density $\rho_{\chi}=1[g/cm^3]$ are used.
%The solid curves include the self-absorption effects and the dotted curves are without self-absorption (No SA).
The top panel displays the cases for dark matter masses $m_{\chi} = 5 \text{MeV}, 50 \text{MeV}, 500 \text{MeV}$ with a magnetic field coefficient \( B_0 = 10^5 \, \text{G} \). The bottom panel illustrates various magnetic field amplitudes with \( B_0 = 10^6, 10^5, 10^4 \, \text{G} \) for a constant dark matter mass of $m_{\chi} = 50 \text{MeV}$.
     %%  In the rest of the paper, we include this self-absorption cutoff in our analysis.
%% Larmor frequency $\nu_L =eB/2\pi m_e =2.8 \times 10^{11}(B/10^5 G) $.
%% SKA is sensitive to $\sim 50$ GHz. (0.05-47.92 GHz in SKA1.txt in my data file) Dark matter mass dependence of $\Gamma^{total} \propto m_{\chi}^{-1}$ and that of the timescale for the radiation cooling $\propto \gamma m_e c^2 \propto m_{\chi} $ cancel out. This leads to the weak dependence of the emitted total energy (integrated over the frequency) on the dark matter mass as long as the timescale for the synchrotron radiation is small enough compared with escape timescale. Once the escape timescale becomes smaller than the radiation cooling timescale, the difference among the dark matter mass becomes prominent.
%%        the white dwarf 100 pc away. For SKA sensitivity, I assumed the band width $\Delta B=0.3 \nu$. The white dwarf radius is R = 0.01*Sunradius. 
%%    The magnetosphere radius is  Rmax = R * 10.0
      %%     $B(r)=B_0 (r/R)^{-3}$, B0 = 1.e8 gauss  at the at white dwarf surface
    }
%%      %%  Made by /Users/kenji/BGD/work/codes/kisakaPulsarRadiation/Nov11Sept15nuSnuDecay.ipynb
%%      %  pdflatex -shell-escape draft1.tex  ->  bibtex draft1 -> pdflatex -shell-escape draft1
     
%%  %   \caption{Common caption for both figures}
    \label{fig1:nov17espectruma}
\end{figure}
%%%%%%%%%%%%%%%%%%%%%%%%%%%%%
%%%%%%%%%%%%%%%%%%%%%%%%%%

Fig. \ref{fig1:nov17espectruma} shows the energy spectrum distribution $\nu S_{\nu}$ of our estimated signals (flux density multiplied by frequency, and the integral $\nu S_{\nu}$ over $\ln \nu$ represents the total energy flux) for different dark matter masses (top panel) and for different magnetic field amplitudes (bottom), along with the SKA sensitivities for $t_{obs}=100$ and $1000$ hours of observations.

Fig. \ref{fig1:nov17espectruma} also illustrates the cutoff at low frequencies, which is a characteristic feature of synchrotron radiation. The synchrotron radiation becomes negligible at frequencies below the Larmor frequency $\nu_L =eB(r)/2\pi m_e$ \cite{1979rpa..book.....R,2013LNP...873.....G}. For illustration purposes, we applied a cutoff to the synchrotron radiation power $P_{\nu}(\nu,r)$ so that the radiation vanishes when the frequency is below the Larmor frequency $\nu\leq \nu_L$ in our integrating the radiation contributions over the white dwarf magnetosphere.

%In the rest of the paper, we include this self-absorption cutoff in our analysis.
%Larmor frequency $\nu_L =eB/2\pi m_e =2.8 \times 10^{11}(B/10^5 G) $.
%Fig. \ref{fig1:nov17espectruma} illustrates the suppression due to the self-absorption at a low frequency and we also show the plots with no self-absorption (No SA) for comparison.

The top panel of Fig. \ref{fig1:nov17espectruma} displays the energy spectrum distribution across different dark matter masses, while the bottom panel illustrates variations due to changing magnetic fields. We observe a general trend of increasing peak frequency and emitted total power (represented by the area under the curve) in the figure with larger dark matter masses and magnetic field strengths. The shift in peak frequency however is not simply proportional to \(\gamma^2 B_0\), and likewise, the total emitted power dependence on $\gamma, B_0$ is not trivial either. This is attributed to the integration of the synchrotron radiation contribution over the position dependent magnetic field profile within the white dwarf magnetosphere. For a larger radius, the magnetic field becomes small even though there is a bigger volume factor. In fact, in some regions of the magnetosphere, particularly for smaller values of \(\gamma\) and \(B(r)\), synchrotron radiation cooling is less efficient and the advection timescale can become comparable to, or shorter than, the radiation cooling timescale. The resultant synchrotron radiation energy spectrum hence can posses non-trivial dependence on $\gamma, B$.

\begin{figure}[!htbp]

     \begin{tabular}{c}
       \includegraphics[width=0.5\textwidth]{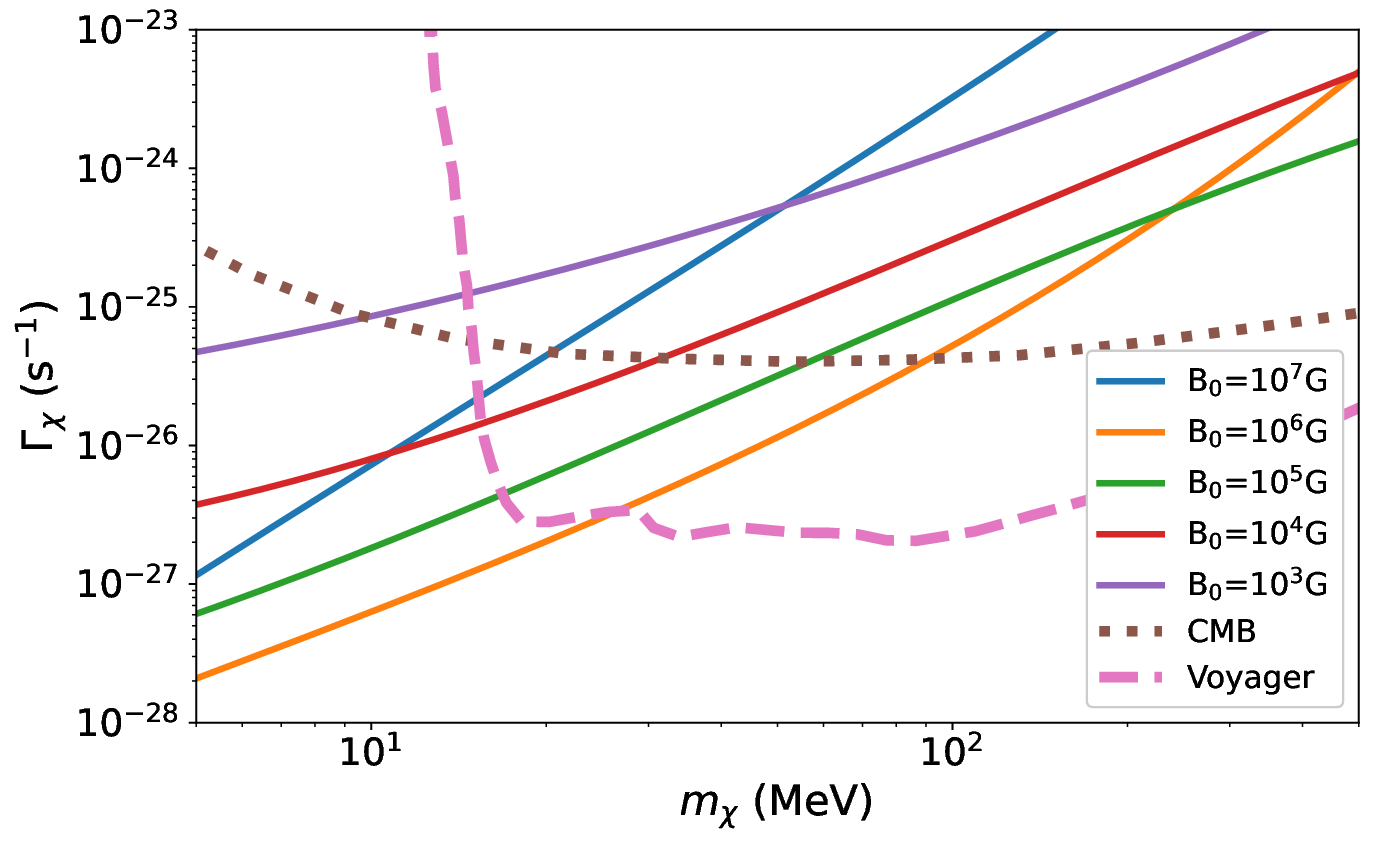}
%                \diracode/taulimitnov11}
     \end{tabular}

     % inkscape input.pdf --export-eps=output.eps  (uses inkscape) <- works best
     % pdftops -eps my.pdf my.eps (uses poppler)
  %\vspace{-25mm}
     \caption{The lower bounds on the dark matter decay rate $\Gamma_{\chi}$ to be detectable by the SKA assuming 100 hours of observation. The dipole magnetic field profile $B(r)=B_0 (r/r_0)^{-3}$ is assumed and the magnetosphere spans from $r_0=0.01 R_{\odot}$ to $10r_0$. The bounds for $B_0=10^3-10^7$G are shown. The constant dark matter density $\rho_{\chi}=1 g/cm^3$ is assumed. The upper bounds from the CMB and Voyager data are also plotted for reference.}
   \label{fig1:taulimit}
\end{figure}

Having obtained the radiation spectrum, we can now put the bounds on the dark matter decay rate $\Gamma_{\chi}$. We conservatively obtain the the bounds on $\Gamma_{\chi}$ by ensuring that the synchrotron radiation flux does not exceed the SKA threshold across the entire frequency the SKA is sensitive to. It is around 50 MHz to 50 GHz as illustrated in our figures, and we assume 100 hours of SKA observation in deriving the bounds.
Fig. \ref{fig1:taulimit} shows the resultant lower bounds on $\Gamma_{\chi}$ above which our synchrotron radiation signals can be detectable by the SKA.
%For SKA sensitivity, the band width $\Delta B=0.3 \nu$ is assumed. %The dark matter mass needs to be at least twice of the electron mass to decay into an electron positron pair. We chose 5 MeV for concreteness in the figure. 
The behavior of the curves for different magnetic fields are caused by the change of the peak frequency and the change of the synchrotron radiation power amplitude. $B_0=10^7$G has a peak frequency well above the SKA sensitive frequencies. Hence the lower magnetic field $B_0=10^6G$ with a lower peak frequency can give tighter bounds even though the radiation power is smaller due to the smaller magnetic field (the total synchrotron radiation power is proportional to $B^2 \gamma^2$). For even lower magnetic fields with $B_0\lesssim 10^5$G, the synchrotron radiation power reduction causes the weaker bounds on $\Gamma_{\chi}$. % The quantitative behavior also depends on the SKA sensitivity dependence on the frequency.
%(also because of the escape timescale becomes shorter than the radiation timescale for smaller $\gamma$ and $B$)
%The power radiated per unit frequency is proportional to $B$ and the total power integrated over the frequency is proportional to $B^2 \gamma^2 $ (the total power determines the timescale for the radiation). the peak frequency is proportional to $B \gamma^2$.
%       The white dwarf 100 pc away. For SKA sensitivity, I assumed the band width $\Delta B=0.3 \nu$. The dark matter mass needs to be at least twice of the electron mass to decay into an electron positron pair. We chose 5 MeV for concreteness in the figure. 
%       The radiation energy $\nu S_{\nu}$ is common for the same dark matter mass.
Besides the requirement for the dark matter to have the life time longer than the age of the Universe $\Gamma_{\chi}^{-1} \gtrsim 4 \times 10^{17} [s]$, there are much tighter bounds from other astrophysical observations. For the sub-GeV dark matter, the CMB and Voyager give among the tightest bounds and the upper bounds from the Planck CMB and Voyager data are plotted in Fig. 3 for reference \cite{Slatyer:2016qyl, Slatyer:2017sev,Liu:2020wqz,Boudaud:2016mos}. Dark matter decays can inject energy into the cosmic plasma, altering the reionization history and affecting the CMB data. The Voyager data can put bounds on the dark matter from the cosmic ray measurements in the interstellar medium outside the influence of the solar wind \footnote{The cosmic ray electrons/positrons are usually shielded by the solar magnetic fields for the earth-bound detectors (so-called solar modulation effects), but not for the Voyager which crossed the heliopause. Such Voyager's unique data sets were compared with a simulation of the expected cosmic-ray electron and positron fluxes originating from dark matter decay to obtain the novel constraints on the dark matter properties \cite{Boudaud:2016mos,Boudaud:2021zzd,Boudaud:2018hqb}. The bounds shown in our figure are from the diffuse DM distributions, which are also applicable to our study.}. Our study demonstrates that our scenarios involving the white dwarf and black hole can potentially lead to the signals observable by the SKA even with a small enough dark matter decay rate satisfying other stringent bounds.
We also note that there are many magnetic white dwarfs closer than 100 pc away \cite{Enoto:2019vcg,Ferrario:2015oda}, and our signals may well be bigger than those in this figure using the fiducial value of 100 pc (the readers can straightforwardly scale the signals which are inversely proportional to the luminosity distance squared).

%% For the sub-GeV dark matter, the CMB gives among the tightest bounds and the upper bound from the Planck CMB data is plotted in Fig. \ref{fig1:taulimit} for reference \cite{Slatyer:2016qyl, Slatyer:2017sev}. Dark matter decays can inject energy into the cosmic plasma, altering the reionization history and affecting the CMB data. Our study demonstrates that our scenarios involving the white dwarf and black hole can potentially lead to the signals observable by the SKA even with a small enough dark matter decay rate satisfying the CMB bounds. 

%%%%%%%%%%%%%%%%%%%%%
%%%%%%%%%%%%%%%%%

 \begin{figure}%[!htbp]

     \begin{tabular}{c}
       \includegraphics[width=0.5\textwidth]{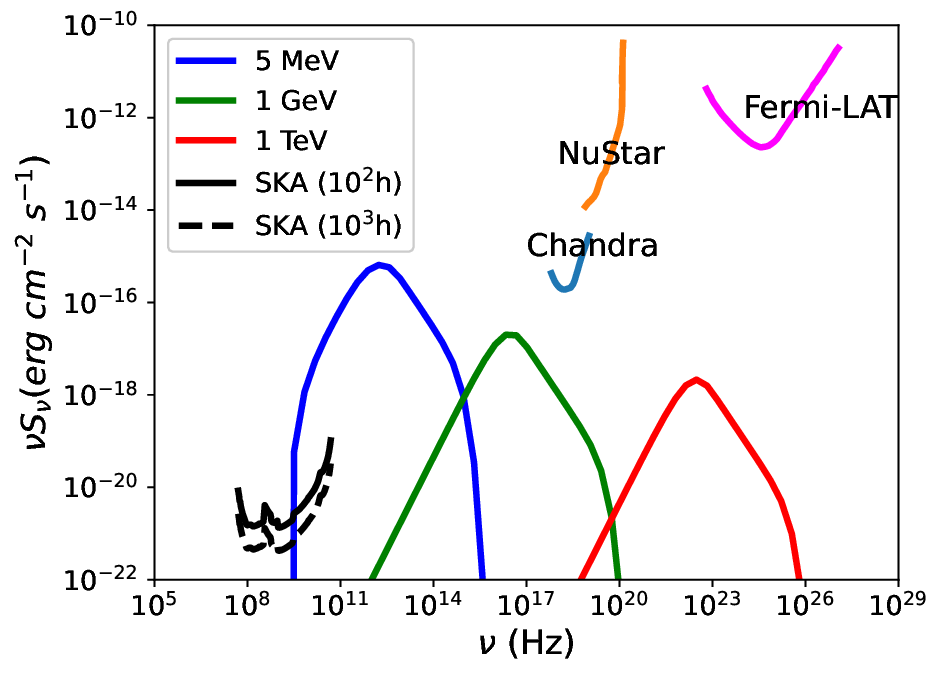} % 1e6 Gauss, self-absorption on
   %             \diracode/2gammaxraylimitNov11}
     \end{tabular}

     % inkscape input.pdf --export-eps=output.eps  (uses inkscape) <- works best
     % pdftops -eps my.pdf my.eps (uses poppler)
  %\vspace{-25mm}
     \caption{
       The energy spectrum distribution $\nu S_{\nu}$ for the synchrotron radiation as a function of the frequency. The magnetic white dwarf is assumed to be 100 pc away and the dark matter density $\rho_{\chi}=1g/cm^3$ is assumed. The magnetic field follows the dipole profile $B(r)=B_0(r/r_0)^{-3}$ with $B_0=10^6G$ and the magnetosphere region is from $r_0=0.01 R_{\odot}$ to $10 r_0$. The dark matter masses for $m_{\chi}=5MeV,1GeV,1TeV$ are shown. The sensitivity of different experiments are also shown covering the radio (SKA assuming 100 and 1000 hours of observations), X-ray (Chandra, NuSTAR) and gamma-ray (Fermi-LAT) frequencies. 100 hours of observations are assumed for Chandra, NuSTAR and Fermi-LAT sensitivity curves.}
   \label{fig:wd1}
\end{figure}
 Before concluding our discussions, let us also briefly discuss the prospects of the synchrotron radiation for our scenarios at the frequencies higher than the radio range relevant for the SKA. For illustration, Fig. \ref{fig:wd1} shows the energy spectrum covering X-ray and gamma-ray frequencies in addition to the radio range. For the large dark matter mass above GeV scale, there currently exist tight upper bounds on dark matter decay rate $\Gamma_{\chi}$ from the X-ray and gamma-ray indirect dark matter search \cite{Slatyer:2017sev}. % or equivalently the lower bounds on the dark matter life time $\tau=\Gamma^{-1}$,
 For instance, for the sub-GeV dark matter, the bounds are of order $\Gamma ^{-1} \gtrsim 10^{25}$ sec, and, for the mass above 1 GeV, the bounds are of order $10^{27}\sim 10^{28}$ sec. We accordingly adopt the fiducial vales of $\Gamma=10^{-25} [s^{-1}]$  for $m_{\chi}<1$ GeV, $10^{-27}[s^{-1}]$ for $1 GeV\leq m_{\chi}<1 TeV$ and $10^{-28}[s^{-1}]$ for $m_{\chi}\geq 1$ TeV in calculating our synchtron radiation signals in Fig. \ref{fig:wd1} \cite{Slatyer:2016qyl,Essig:2013goa,Cohen:2016uyg}. Fig. \ref{fig:wd1} also shows the sensitivity of the currently available observation data for the X-ray and gamma-ray observations (Chandra, NuSTAR and Fermi assuming 100 hours of observations) \cite{Cembranos:2019noa,2005SPIE.5900..266K,2002sf2a.conf..271F,Lucchetta:2022nrm}.
We can see that experiments targeting frequencies higher than the radio range exhibit less competitive sensitivity compared to the superior sensitivity expected from the SKA. This is partly due to the large effective area that radio telescopes, such as the SKA, can boast.
%% Let us also mention that the magnetic white dwarf is preferable for our scenarios than the neutron stars. Even though the neutron star tends to have a bigger magnetic field than the white dwarf and hence the amplitude of the radiation power bigger, the too high magnetic field is disadvantageous for us because the peak frequency can be too high for the SKA sensitive frequency range ($50M\sim 50 G$Hz).
%% Furthermore, the larger volume of white dwarf magnetosphere compared to that of a neutron star bolsters our preference for white dwarfs in our scenarios.
\\
\\
We investigated potential SKA probes of synchrotron radiation resulting from sub-GeV dark matter decay in the presence of strong magnetic fields. We specifically examined scenarios where dark matter environments around a primordial black hole overlap with the magnetosphere of a magnetic white dwarf. Our estimates suggest that the encounters of compact objects, such as white dwarfs and black holes, could present compelling targets for upcoming radio telescopes to clarify the nature of dark matter. While our scenarios primarily require the singular encounters with unbound orbits, the possibly rarer yet intriguing scenarios of binary formation would also warrant further exploration.
The black hole-white dwarf binary system has been less explored in the literature compared with black hole-black hole and black hole-neutron star binaries. The black hole-white dwarf binaries can be of great interest for the targets of multi-messenger objects. The black hole-white dwarf binaries have been also studied for the gravitational wave signals. For example, the detection of ${\cal O}(10)$ PBH-White dwarf merger events is feasible for $M_{PBH}\sim 10^5 M_{\odot}$ by the DECIGO-like space-borne gravitational wave interferometer within three years at distances up to \(\sim 1\) Gpc scale \cite{Yamamoto:2023tsr}. %even though an  can be possible by the PBH clustering and while the event rate decreases (increases) for a smaller (larger) mass (for instance, for $M_{PBH}\sim 1 M_{\odot}$, ${\cal O}(10)$ event detection is feasible) \cite{Yamamoto:2023tsr}.
We however note that these estimations of event rates significantly depend on the assumptions of underlying models, such as the mass functions of compact objects. For more detailed discussions on calculation methodologies and model dependencies for the estimation of the compact object encountering rate, we direct readers to the pertinent literature \cite{Yamamoto:2023tsr, Nelemans:2001hp, Maguire:2020lad,Wang:2020jsx,Xuan:2022qkw, 2008MNRAS.391..718S, LISA:2022yao, Shao:2021dbg, Qin:2023tkb, 2009ApJ...695..404R}.

The detailed estimation for the black hole-white dwarf encountering rates relevant for our scenarios are left for future work. The potential X-ray, gamma-ray and gravitational wave signals have been studied for the white dwarf-black hole binaries, but radio signals have received little attention \cite{Maguire:2020lad,2009ApJ...695..404R,2008MNRAS.391..718S,LISA:2022yao,2023MNRAS.tmp.3162Y,Nelemans:2001hp,LISA:2022yao,Nelemans:2001hp,Qin:2023tkb, Ye:2023fpb,Shao:2021dbg}. More detailed analyses of our scenarios, including tidal effects and the possible binary formation, will be addressed in future numerical studies.

\section*{Acknowledgment}
We thank N. Kitajima and K. Toma for the useful discussions. KK thanks the CTPU at IBS for their hospitality during the completion of this work. This work was supported by Center of Quantum Cosmo Theoretical Physics (NSFC grant No. 12347103), KAKENHI Grants No. JP21H01078, No. JP22H01267, and No. JP22K03681.

\bibliography{/Users/kenji/CGD/work/paper/kenjireference}

\begin{thebibliography}{108}%
\makeatletter
\providecommand \@ifxundefined [1]{%
 \@ifx{#1\undefined}
}%
\providecommand \@ifnum [1]{%
 \ifnum #1\expandafter \@firstoftwo
 \else \expandafter \@secondoftwo
 \fi
}%
\providecommand \@ifx [1]{%
 \ifx #1\expandafter \@firstoftwo
 \else \expandafter \@secondoftwo
 \fi
}%
\providecommand \natexlab [1]{#1}%
\providecommand \enquote  [1]{``#1''}%
\providecommand \bibnamefont  [1]{#1}%
\providecommand \bibfnamefont [1]{#1}%
\providecommand \citenamefont [1]{#1}%
\providecommand \href@noop [0]{\@secondoftwo}%
\providecommand \href [0]{\begingroup \@sanitize@url \@href}%
\providecommand \@href[1]{\@@startlink{#1}\@@href}%
\providecommand \@@href[1]{\endgroup#1\@@endlink}%
\providecommand \@sanitize@url [0]{\catcode `\\12\catcode `\$12\catcode
  `\&12\catcode `\#12\catcode `\^12\catcode `\_12\catcode `\%12\relax}%
\providecommand \@@startlink[1]{}%
\providecommand \@@endlink[0]{}%
\providecommand \url  [0]{\begingroup\@sanitize@url \@url }%
\providecommand \@url [1]{\endgroup\@href {#1}{\urlprefix }}%
\providecommand \urlprefix  [0]{URL }%
\providecommand \Eprint [0]{\href }%
\providecommand \doibase [0]{https://doi.org/}%
\providecommand \selectlanguage [0]{\@gobble}%
\providecommand \bibinfo  [0]{\@secondoftwo}%
\providecommand \bibfield  [0]{\@secondoftwo}%
\providecommand \translation [1]{[#1]}%
\providecommand \BibitemOpen [0]{}%
\providecommand \bibitemStop [0]{}%
\providecommand \bibitemNoStop [0]{.\EOS\space}%
\providecommand \EOS [0]{\spacefactor3000\relax}%
\providecommand \BibitemShut  [1]{\csname bibitem#1\endcsname}%
\let\auto@bib@innerbib\@empty
%</preamble>
\bibitem [{\citenamefont {Jungman}\ \emph {et~al.}(1996)\citenamefont
  {Jungman}, \citenamefont {Kamionkowski},\ and\ \citenamefont
  {Griest}}]{Jungman:1995df}%
  \BibitemOpen
  \bibfield  {author} {\bibinfo {author} {\bibfnamefont {G.}~\bibnamefont
  {Jungman}}, \bibinfo {author} {\bibfnamefont {M.}~\bibnamefont
  {Kamionkowski}},\ and\ \bibinfo {author} {\bibfnamefont {K.}~\bibnamefont
  {Griest}},\ }\bibfield  {title} {\bibinfo {title} {{Supersymmetric dark
  matter}},\ }\href {https://doi.org/10.1016/0370-1573(95)00058-5} {\bibfield
  {journal} {\bibinfo  {journal} {Phys. Rept.}\ }\textbf {\bibinfo {volume}
  {267}},\ \bibinfo {pages} {195} (\bibinfo {year} {1996})},\ \Eprint
  {https://arxiv.org/abs/hep-ph/9506380} {arXiv:hep-ph/9506380} \BibitemShut
  {NoStop}%
\bibitem [{\citenamefont {Arcadi}\ \emph {et~al.}(2018)\citenamefont {Arcadi},
  \citenamefont {Dutra}, \citenamefont {Ghosh}, \citenamefont {Lindner},
  \citenamefont {Mambrini}, \citenamefont {Pierre}, \citenamefont {Profumo},\
  and\ \citenamefont {Queiroz}}]{Arcadi:2017kky}%
  \BibitemOpen
  \bibfield  {author} {\bibinfo {author} {\bibfnamefont {G.}~\bibnamefont
  {Arcadi}}, \bibinfo {author} {\bibfnamefont {M.}~\bibnamefont {Dutra}},
  \bibinfo {author} {\bibfnamefont {P.}~\bibnamefont {Ghosh}}, \bibinfo
  {author} {\bibfnamefont {M.}~\bibnamefont {Lindner}}, \bibinfo {author}
  {\bibfnamefont {Y.}~\bibnamefont {Mambrini}}, \bibinfo {author}
  {\bibfnamefont {M.}~\bibnamefont {Pierre}}, \bibinfo {author} {\bibfnamefont
  {S.}~\bibnamefont {Profumo}},\ and\ \bibinfo {author} {\bibfnamefont {F.~S.}\
  \bibnamefont {Queiroz}},\ }\bibfield  {title} {\bibinfo {title} {{The waning
  of the WIMP? A review of models, searches, and constraints}},\ }\href
  {https://doi.org/10.1140/epjc/s10052-018-5662-y} {\bibfield  {journal}
  {\bibinfo  {journal} {Eur. Phys. J. C}\ }\textbf {\bibinfo {volume} {78}},\
  \bibinfo {pages} {203} (\bibinfo {year} {2018})},\ \Eprint
  {https://arxiv.org/abs/1703.07364} {arXiv:1703.07364 [hep-ph]} \BibitemShut
  {NoStop}%
\bibitem [{\citenamefont {Boddy}\ \emph {et~al.}(2022)\citenamefont {Boddy}
  \emph {et~al.}}]{Boddy:2022knd}%
  \BibitemOpen
  \bibfield  {author} {\bibinfo {author} {\bibfnamefont {K.~K.}\ \bibnamefont
  {Boddy}} \emph {et~al.},\ }\bibfield  {title} {\bibinfo {title}
  {{Snowmass2021 theory frontier white paper: Astrophysical and cosmological
  probes of dark matter}},\ }\href
  {https://doi.org/10.1016/j.jheap.2022.06.005} {\bibfield  {journal} {\bibinfo
   {journal} {JHEAp}\ }\textbf {\bibinfo {volume} {35}},\ \bibinfo {pages}
  {112} (\bibinfo {year} {2022})},\ \Eprint {https://arxiv.org/abs/2203.06380}
  {arXiv:2203.06380 [hep-ph]} \BibitemShut {NoStop}%
\bibitem [{\citenamefont {Chou}\ \emph {et~al.}(2022)\citenamefont {Chou} \emph
  {et~al.}}]{Chou:2022luk}%
  \BibitemOpen
  \bibfield  {author} {\bibinfo {author} {\bibfnamefont {A.~S.}\ \bibnamefont
  {Chou}} \emph {et~al.},\ }\bibfield  {title} {\bibinfo {title} {{Snowmass
  Cosmic Frontier Report}},\ }in\ \href@noop {} {\emph {\bibinfo {booktitle}
  {{2022 Snowmass Summer Study}}}}\ (\bibinfo {year} {2022})\ \Eprint
  {https://arxiv.org/abs/2211.09978} {arXiv:2211.09978 [hep-ex]} \BibitemShut
  {NoStop}%
\bibitem [{\citenamefont {Carr}\ and\ \citenamefont
  {Kuhnel}(2020)}]{Carr:2020xqk}%
  \BibitemOpen
  \bibfield  {author} {\bibinfo {author} {\bibfnamefont {B.}~\bibnamefont
  {Carr}}\ and\ \bibinfo {author} {\bibfnamefont {F.}~\bibnamefont {Kuhnel}},\
  }\bibfield  {title} {\bibinfo {title} {{Primordial Black Holes as Dark
  Matter: Recent Developments}},\ }\href@noop {} {\  (\bibinfo {year}
  {2020})},\ \Eprint {https://arxiv.org/abs/2006.02838} {arXiv:2006.02838
  [astro-ph.CO]} \BibitemShut {NoStop}%
\bibitem [{\citenamefont {Chadha-Day}\ \emph {et~al.}(2022)\citenamefont
  {Chadha-Day}, \citenamefont {Ellis},\ and\ \citenamefont
  {Marsh}}]{Chadha-Day:2021szb}%
  \BibitemOpen
  \bibfield  {author} {\bibinfo {author} {\bibfnamefont {F.}~\bibnamefont
  {Chadha-Day}}, \bibinfo {author} {\bibfnamefont {J.}~\bibnamefont {Ellis}},\
  and\ \bibinfo {author} {\bibfnamefont {D.~J.~E.}\ \bibnamefont {Marsh}},\
  }\bibfield  {title} {\bibinfo {title} {{Axion dark matter: What is it and why
  now?}},\ }\href {https://doi.org/10.1126/sciadv.abj3618} {\bibfield
  {journal} {\bibinfo  {journal} {Sci. Adv.}\ }\textbf {\bibinfo {volume}
  {8}},\ \bibinfo {pages} {abj3618} (\bibinfo {year} {2022})},\ \Eprint
  {https://arxiv.org/abs/2105.01406} {arXiv:2105.01406 [hep-ph]} \BibitemShut
  {NoStop}%
\bibitem [{\citenamefont {Knapen}\ \emph {et~al.}(2017)\citenamefont {Knapen},
  \citenamefont {Lin},\ and\ \citenamefont {Zurek}}]{Knapen:2017xzo}%
  \BibitemOpen
  \bibfield  {author} {\bibinfo {author} {\bibfnamefont {S.}~\bibnamefont
  {Knapen}}, \bibinfo {author} {\bibfnamefont {T.}~\bibnamefont {Lin}},\ and\
  \bibinfo {author} {\bibfnamefont {K.~M.}\ \bibnamefont {Zurek}},\ }\bibfield
  {title} {\bibinfo {title} {{Light Dark Matter: Models and Constraints}},\
  }\href {https://doi.org/10.1103/PhysRevD.96.115021} {\bibfield  {journal}
  {\bibinfo  {journal} {Phys. Rev. D}\ }\textbf {\bibinfo {volume} {96}},\
  \bibinfo {pages} {115021} (\bibinfo {year} {2017})},\ \Eprint
  {https://arxiv.org/abs/1709.07882} {arXiv:1709.07882 [hep-ph]} \BibitemShut
  {NoStop}%
\bibitem [{\citenamefont {Battaglieri}\ \emph {et~al.}(2017)\citenamefont
  {Battaglieri} \emph {et~al.}}]{Battaglieri:2017aum}%
  \BibitemOpen
  \bibfield  {author} {\bibinfo {author} {\bibfnamefont {M.}~\bibnamefont
  {Battaglieri}} \emph {et~al.},\ }\bibfield  {title} {\bibinfo {title} {{US
  Cosmic Visions: New Ideas in Dark Matter 2017: Community Report}},\ }in\
  \href@noop {} {\emph {\bibinfo {booktitle} {{U.S. Cosmic Visions: New Ideas
  in Dark Matter}}}}\ (\bibinfo {year} {2017})\ \Eprint
  {https://arxiv.org/abs/1707.04591} {arXiv:1707.04591 [hep-ph]} \BibitemShut
  {NoStop}%
\bibitem [{\citenamefont {Bondarenko}\ \emph {et~al.}(2020)\citenamefont
  {Bondarenko}, \citenamefont {Boyarsky}, \citenamefont {Bringmann},
  \citenamefont {Hufnagel}, \citenamefont {Schmidt-Hoberg},\ and\ \citenamefont
  {Sokolenko}}]{Bondarenko:2019vrb}%
  \BibitemOpen
  \bibfield  {author} {\bibinfo {author} {\bibfnamefont {K.}~\bibnamefont
  {Bondarenko}}, \bibinfo {author} {\bibfnamefont {A.}~\bibnamefont
  {Boyarsky}}, \bibinfo {author} {\bibfnamefont {T.}~\bibnamefont {Bringmann}},
  \bibinfo {author} {\bibfnamefont {M.}~\bibnamefont {Hufnagel}}, \bibinfo
  {author} {\bibfnamefont {K.}~\bibnamefont {Schmidt-Hoberg}},\ and\ \bibinfo
  {author} {\bibfnamefont {A.}~\bibnamefont {Sokolenko}},\ }\bibfield  {title}
  {\bibinfo {title} {{Direct detection and complementary constraints for
  sub-GeV dark matter}},\ }\href {https://doi.org/10.1007/JHEP03(2020)118}
  {\bibfield  {journal} {\bibinfo  {journal} {JHEP}\ }\textbf {\bibinfo
  {volume} {03}},\ \bibinfo {pages} {118}},\ \Eprint
  {https://arxiv.org/abs/1909.08632} {arXiv:1909.08632 [hep-ph]} \BibitemShut
  {NoStop}%
\bibitem [{\citenamefont {Coogan}\ \emph {et~al.}(2020)\citenamefont {Coogan},
  \citenamefont {Morrison},\ and\ \citenamefont {Profumo}}]{Coogan:2019qpu}%
  \BibitemOpen
  \bibfield  {author} {\bibinfo {author} {\bibfnamefont {A.}~\bibnamefont
  {Coogan}}, \bibinfo {author} {\bibfnamefont {L.}~\bibnamefont {Morrison}},\
  and\ \bibinfo {author} {\bibfnamefont {S.}~\bibnamefont {Profumo}},\
  }\bibfield  {title} {\bibinfo {title} {{Hazma: A Python Toolkit for Studying
  Indirect Detection of Sub-GeV Dark Matter}},\ }\href
  {https://doi.org/10.1088/1475-7516/2020/01/056} {\bibfield  {journal}
  {\bibinfo  {journal} {JCAP}\ }\textbf {\bibinfo {volume} {01}},\ \bibinfo
  {pages} {056}},\ \Eprint {https://arxiv.org/abs/1907.11846} {arXiv:1907.11846
  [hep-ph]} \BibitemShut {NoStop}%
\bibitem [{\citenamefont {Essig}\ \emph {et~al.}(2013)\citenamefont {Essig},
  \citenamefont {Kuflik}, \citenamefont {McDermott}, \citenamefont {Volansky},\
  and\ \citenamefont {Zurek}}]{Essig:2013goa}%
  \BibitemOpen
  \bibfield  {author} {\bibinfo {author} {\bibfnamefont {R.}~\bibnamefont
  {Essig}}, \bibinfo {author} {\bibfnamefont {E.}~\bibnamefont {Kuflik}},
  \bibinfo {author} {\bibfnamefont {S.~D.}\ \bibnamefont {McDermott}}, \bibinfo
  {author} {\bibfnamefont {T.}~\bibnamefont {Volansky}},\ and\ \bibinfo
  {author} {\bibfnamefont {K.~M.}\ \bibnamefont {Zurek}},\ }\bibfield  {title}
  {\bibinfo {title} {{Constraining Light Dark Matter with Diffuse X-Ray and
  Gamma-Ray Observations}},\ }\href {https://doi.org/10.1007/JHEP11(2013)193}
  {\bibfield  {journal} {\bibinfo  {journal} {JHEP}\ }\textbf {\bibinfo
  {volume} {11}},\ \bibinfo {pages} {193}},\ \Eprint
  {https://arxiv.org/abs/1309.4091} {arXiv:1309.4091 [hep-ph]} \BibitemShut
  {NoStop}%
\bibitem [{\citenamefont {Cirelli}\ \emph {et~al.}(2021)\citenamefont
  {Cirelli}, \citenamefont {Fornengo}, \citenamefont {Kavanagh},\ and\
  \citenamefont {Pinetti}}]{Cirelli:2020bpc}%
  \BibitemOpen
  \bibfield  {author} {\bibinfo {author} {\bibfnamefont {M.}~\bibnamefont
  {Cirelli}}, \bibinfo {author} {\bibfnamefont {N.}~\bibnamefont {Fornengo}},
  \bibinfo {author} {\bibfnamefont {B.~J.}\ \bibnamefont {Kavanagh}},\ and\
  \bibinfo {author} {\bibfnamefont {E.}~\bibnamefont {Pinetti}},\ }\bibfield
  {title} {\bibinfo {title} {{Integral X-ray constraints on sub-GeV Dark
  Matter}},\ }\href {https://doi.org/10.1103/PhysRevD.103.063022} {\bibfield
  {journal} {\bibinfo  {journal} {Phys. Rev. D}\ }\textbf {\bibinfo {volume}
  {103}},\ \bibinfo {pages} {063022} (\bibinfo {year} {2021})},\ \Eprint
  {https://arxiv.org/abs/2007.11493} {arXiv:2007.11493 [hep-ph]} \BibitemShut
  {NoStop}%
\bibitem [{\citenamefont {Choudhury}\ and\ \citenamefont
  {Sachdeva}(2019)}]{Choudhury:2019tss}%
  \BibitemOpen
  \bibfield  {author} {\bibinfo {author} {\bibfnamefont {D.}~\bibnamefont
  {Choudhury}}\ and\ \bibinfo {author} {\bibfnamefont {D.}~\bibnamefont
  {Sachdeva}},\ }\bibfield  {title} {\bibinfo {title} {{Model independent
  analysis of MeV scale dark matter: Cosmological constraints}},\ }\href
  {https://doi.org/10.1103/PhysRevD.100.035007} {\bibfield  {journal} {\bibinfo
   {journal} {Phys. Rev. D}\ }\textbf {\bibinfo {volume} {100}},\ \bibinfo
  {pages} {035007} (\bibinfo {year} {2019})},\ \Eprint
  {https://arxiv.org/abs/1903.06049} {arXiv:1903.06049 [hep-ph]} \BibitemShut
  {NoStop}%
\bibitem [{\citenamefont {Hooper}\ \emph {et~al.}(2007)\citenamefont {Hooper},
  \citenamefont {Kaplinghat}, \citenamefont {Strigari},\ and\ \citenamefont
  {Zurek}}]{Hooper:2007tu}%
  \BibitemOpen
  \bibfield  {author} {\bibinfo {author} {\bibfnamefont {D.}~\bibnamefont
  {Hooper}}, \bibinfo {author} {\bibfnamefont {M.}~\bibnamefont {Kaplinghat}},
  \bibinfo {author} {\bibfnamefont {L.~E.}\ \bibnamefont {Strigari}},\ and\
  \bibinfo {author} {\bibfnamefont {K.~M.}\ \bibnamefont {Zurek}},\ }\bibfield
  {title} {\bibinfo {title} {{MeV Dark Matter and Small Scale Structure}},\
  }\href {https://doi.org/10.1103/PhysRevD.76.103515} {\bibfield  {journal}
  {\bibinfo  {journal} {Phys. Rev. D}\ }\textbf {\bibinfo {volume} {76}},\
  \bibinfo {pages} {103515} (\bibinfo {year} {2007})},\ \Eprint
  {https://arxiv.org/abs/0704.2558} {arXiv:0704.2558 [astro-ph]} \BibitemShut
  {NoStop}%
\bibitem [{\citenamefont {Tashiro}\ \emph {et~al.}(2014)\citenamefont
  {Tashiro}, \citenamefont {Kadota},\ and\ \citenamefont
  {Silk}}]{Tashiro:2014tsa}%
  \BibitemOpen
  \bibfield  {author} {\bibinfo {author} {\bibfnamefont {H.}~\bibnamefont
  {Tashiro}}, \bibinfo {author} {\bibfnamefont {K.}~\bibnamefont {Kadota}},\
  and\ \bibinfo {author} {\bibfnamefont {J.}~\bibnamefont {Silk}},\ }\bibfield
  {title} {\bibinfo {title} {{Effects of dark matter-baryon scattering on
  redshifted 21 cm signals}},\ }\href
  {https://doi.org/10.1103/PhysRevD.90.083522} {\bibfield  {journal} {\bibinfo
  {journal} {Phys. Rev. D}\ }\textbf {\bibinfo {volume} {90}},\ \bibinfo
  {pages} {083522} (\bibinfo {year} {2014})},\ \Eprint
  {https://arxiv.org/abs/1408.2571} {arXiv:1408.2571 [astro-ph.CO]}
  \BibitemShut {NoStop}%
\bibitem [{\citenamefont {Xu}\ \emph {et~al.}(2018)\citenamefont {Xu},
  \citenamefont {Dvorkin},\ and\ \citenamefont {Chael}}]{Xu:2018efh}%
  \BibitemOpen
  \bibfield  {author} {\bibinfo {author} {\bibfnamefont {W.~L.}\ \bibnamefont
  {Xu}}, \bibinfo {author} {\bibfnamefont {C.}~\bibnamefont {Dvorkin}},\ and\
  \bibinfo {author} {\bibfnamefont {A.}~\bibnamefont {Chael}},\ }\bibfield
  {title} {\bibinfo {title} {{Probing sub-GeV Dark Matter-Baryon Scattering
  with Cosmological Observables}},\ }\href
  {https://doi.org/10.1103/PhysRevD.97.103530} {\bibfield  {journal} {\bibinfo
  {journal} {Phys. Rev. D}\ }\textbf {\bibinfo {volume} {97}},\ \bibinfo
  {pages} {103530} (\bibinfo {year} {2018})},\ \Eprint
  {https://arxiv.org/abs/1802.06788} {arXiv:1802.06788 [astro-ph.CO]}
  \BibitemShut {NoStop}%
\bibitem [{\citenamefont {Ooba}\ \emph {et~al.}(2019)\citenamefont {Ooba},
  \citenamefont {Tashiro},\ and\ \citenamefont {Kadota}}]{Ooba:2019erm}%
  \BibitemOpen
  \bibfield  {author} {\bibinfo {author} {\bibfnamefont {J.}~\bibnamefont
  {Ooba}}, \bibinfo {author} {\bibfnamefont {H.}~\bibnamefont {Tashiro}},\ and\
  \bibinfo {author} {\bibfnamefont {K.}~\bibnamefont {Kadota}},\ }\bibfield
  {title} {\bibinfo {title} {{Cosmological constraints on the
  velocity-dependent baryon-dark matter coupling}},\ }\href
  {https://doi.org/10.1088/1475-7516/2019/09/020} {\bibfield  {journal}
  {\bibinfo  {journal} {JCAP}\ }\textbf {\bibinfo {volume} {09}},\ \bibinfo
  {pages} {020}},\ \Eprint {https://arxiv.org/abs/1902.00826} {arXiv:1902.00826
  [astro-ph.CO]} \BibitemShut {NoStop}%
\bibitem [{\citenamefont {Boehm}\ \emph {et~al.}(2004)\citenamefont {Boehm},
  \citenamefont {Ensslin},\ and\ \citenamefont {Silk}}]{Boehm:2002yz}%
  \BibitemOpen
  \bibfield  {author} {\bibinfo {author} {\bibfnamefont {C.}~\bibnamefont
  {Boehm}}, \bibinfo {author} {\bibfnamefont {T.~A.}\ \bibnamefont {Ensslin}},\
  and\ \bibinfo {author} {\bibfnamefont {J.}~\bibnamefont {Silk}},\ }\bibfield
  {title} {\bibinfo {title} {{Can Annihilating dark matter be lighter than a
  few GeVs?}},\ }\href {https://doi.org/10.1088/0954-3899/30/3/004} {\bibfield
  {journal} {\bibinfo  {journal} {J. Phys. G}\ }\textbf {\bibinfo {volume}
  {30}},\ \bibinfo {pages} {279} (\bibinfo {year} {2004})},\ \Eprint
  {https://arxiv.org/abs/astro-ph/0208458} {arXiv:astro-ph/0208458}
  \BibitemShut {NoStop}%
\bibitem [{\citenamefont {Cirelli}\ \emph {et~al.}(2023)\citenamefont
  {Cirelli}, \citenamefont {Fornengo}, \citenamefont {Koechler}, \citenamefont
  {Pinetti},\ and\ \citenamefont {Roach}}]{Cirelli:2023tnx}%
  \BibitemOpen
  \bibfield  {author} {\bibinfo {author} {\bibfnamefont {M.}~\bibnamefont
  {Cirelli}}, \bibinfo {author} {\bibfnamefont {N.}~\bibnamefont {Fornengo}},
  \bibinfo {author} {\bibfnamefont {J.}~\bibnamefont {Koechler}}, \bibinfo
  {author} {\bibfnamefont {E.}~\bibnamefont {Pinetti}},\ and\ \bibinfo {author}
  {\bibfnamefont {B.~M.}\ \bibnamefont {Roach}},\ }\bibfield  {title} {\bibinfo
  {title} {{Putting all the X in one basket: Updated X-ray constraints on
  sub-GeV Dark Matter}},\ }\href
  {https://doi.org/10.1088/1475-7516/2023/07/026} {\bibfield  {journal}
  {\bibinfo  {journal} {JCAP}\ }\textbf {\bibinfo {volume} {07}},\ \bibinfo
  {pages} {026}},\ \Eprint {https://arxiv.org/abs/2303.08854} {arXiv:2303.08854
  [hep-ph]} \BibitemShut {NoStop}%
\bibitem [{\citenamefont {Carr}\ \emph {et~al.}(2023)\citenamefont {Carr},
  \citenamefont {Clesse}, \citenamefont {Garcia-Bellido}, \citenamefont
  {Hawkins},\ and\ \citenamefont {Kuhnel}}]{Carr:2023tpt}%
  \BibitemOpen
  \bibfield  {author} {\bibinfo {author} {\bibfnamefont {B.}~\bibnamefont
  {Carr}}, \bibinfo {author} {\bibfnamefont {S.}~\bibnamefont {Clesse}},
  \bibinfo {author} {\bibfnamefont {J.}~\bibnamefont {Garcia-Bellido}},
  \bibinfo {author} {\bibfnamefont {M.}~\bibnamefont {Hawkins}},\ and\ \bibinfo
  {author} {\bibfnamefont {F.}~\bibnamefont {Kuhnel}},\ }\bibfield  {title}
  {\bibinfo {title} {{Observational Evidence for Primordial Black Holes: A
  Positivist Perspective}},\ }\href@noop {} {\  (\bibinfo {year} {2023})},\
  \Eprint {https://arxiv.org/abs/2306.03903} {arXiv:2306.03903 [astro-ph.CO]}
  \BibitemShut {NoStop}%
\bibitem [{\citenamefont {Freese}\ \emph {et~al.}(2022)\citenamefont {Freese},
  \citenamefont {Galstyan}, \citenamefont {Sandick},\ and\ \citenamefont
  {Stengel}}]{Freese:2022ouh}%
  \BibitemOpen
  \bibfield  {author} {\bibinfo {author} {\bibfnamefont {K.}~\bibnamefont
  {Freese}}, \bibinfo {author} {\bibfnamefont {I.}~\bibnamefont {Galstyan}},
  \bibinfo {author} {\bibfnamefont {P.}~\bibnamefont {Sandick}},\ and\ \bibinfo
  {author} {\bibfnamefont {P.}~\bibnamefont {Stengel}},\ }\bibfield  {title}
  {\bibinfo {title} {{Neutrino point source searches for dark matter spikes}},\
  }\href {https://doi.org/10.1088/1475-7516/2022/08/065} {\bibfield  {journal}
  {\bibinfo  {journal} {JCAP}\ }\textbf {\bibinfo {volume} {08}},\ \bibinfo
  {pages} {065}},\ \Eprint {https://arxiv.org/abs/2202.01126} {arXiv:2202.01126
  [astro-ph.CO]} \BibitemShut {NoStop}%
\bibitem [{\citenamefont {Kuhnel}\ and\ \citenamefont
  {Ohlsson}(2019)}]{Kuhnel:2018kwf}%
  \BibitemOpen
  \bibfield  {author} {\bibinfo {author} {\bibfnamefont {F.}~\bibnamefont
  {Kuhnel}}\ and\ \bibinfo {author} {\bibfnamefont {T.}~\bibnamefont
  {Ohlsson}},\ }\bibfield  {title} {\bibinfo {title} {{Decaying Dark Matter in
  Halos of Primordial Black Holes}},\ }\href
  {https://doi.org/10.1140/epjc/s10052-019-7173-x} {\bibfield  {journal}
  {\bibinfo  {journal} {Eur. Phys. J. C}\ }\textbf {\bibinfo {volume} {79}},\
  \bibinfo {pages} {687} (\bibinfo {year} {2019})},\ \Eprint
  {https://arxiv.org/abs/1811.05810} {arXiv:1811.05810 [astro-ph.CO]}
  \BibitemShut {NoStop}%
\bibitem [{\citenamefont {Scott}\ and\ \citenamefont
  {Sivertsson}(2009)}]{Scott:2009tu}%
  \BibitemOpen
  \bibfield  {author} {\bibinfo {author} {\bibfnamefont {P.}~\bibnamefont
  {Scott}}\ and\ \bibinfo {author} {\bibfnamefont {S.}~\bibnamefont
  {Sivertsson}},\ }\bibfield  {title} {\bibinfo {title} {{Gamma-Rays from
  Ultracompact Primordial Dark Matter Minihalos}},\ }\href
  {https://doi.org/10.1103/PhysRevLett.103.211301} {\bibfield  {journal}
  {\bibinfo  {journal} {Phys. Rev. Lett.}\ }\textbf {\bibinfo {volume} {103}},\
  \bibinfo {pages} {211301} (\bibinfo {year} {2009})},\ \bibinfo {note}
  {[Erratum: Phys.Rev.Lett. 105, 119902 (2010)]},\ \Eprint
  {https://arxiv.org/abs/0908.4082} {arXiv:0908.4082 [astro-ph.CO]}
  \BibitemShut {NoStop}%
\bibitem [{\citenamefont {Gondolo}\ and\ \citenamefont
  {Silk}(1999)}]{Gondolo:1999ef}%
  \BibitemOpen
  \bibfield  {author} {\bibinfo {author} {\bibfnamefont {P.}~\bibnamefont
  {Gondolo}}\ and\ \bibinfo {author} {\bibfnamefont {J.}~\bibnamefont {Silk}},\
  }\bibfield  {title} {\bibinfo {title} {{Dark matter annihilation at the
  galactic center}},\ }\href {https://doi.org/10.1103/PhysRevLett.83.1719}
  {\bibfield  {journal} {\bibinfo  {journal} {Phys. Rev. Lett.}\ }\textbf
  {\bibinfo {volume} {83}},\ \bibinfo {pages} {1719} (\bibinfo {year}
  {1999})},\ \Eprint {https://arxiv.org/abs/astro-ph/9906391}
  {arXiv:astro-ph/9906391} \BibitemShut {NoStop}%
\bibitem [{\citenamefont {Lacki}\ and\ \citenamefont
  {Beacom}(2010)}]{Lacki:2010zf}%
  \BibitemOpen
  \bibfield  {author} {\bibinfo {author} {\bibfnamefont {B.~C.}\ \bibnamefont
  {Lacki}}\ and\ \bibinfo {author} {\bibfnamefont {J.~F.}\ \bibnamefont
  {Beacom}},\ }\bibfield  {title} {\bibinfo {title} {{Primordial Black Holes as
  Dark Matter: Almost All or Almost Nothing}},\ }\href
  {https://doi.org/10.1088/2041-8205/720/1/L67} {\bibfield  {journal} {\bibinfo
   {journal} {Astrophys. J. Lett.}\ }\textbf {\bibinfo {volume} {720}},\
  \bibinfo {pages} {L67} (\bibinfo {year} {2010})},\ \Eprint
  {https://arxiv.org/abs/1003.3466} {arXiv:1003.3466 [astro-ph.CO]}
  \BibitemShut {NoStop}%
\bibitem [{\citenamefont {Adamek}\ \emph {et~al.}(2019)\citenamefont {Adamek},
  \citenamefont {Byrnes}, \citenamefont {Gosenca},\ and\ \citenamefont
  {Hotchkiss}}]{Adamek:2019gns}%
  \BibitemOpen
  \bibfield  {author} {\bibinfo {author} {\bibfnamefont {J.}~\bibnamefont
  {Adamek}}, \bibinfo {author} {\bibfnamefont {C.~T.}\ \bibnamefont {Byrnes}},
  \bibinfo {author} {\bibfnamefont {M.}~\bibnamefont {Gosenca}},\ and\ \bibinfo
  {author} {\bibfnamefont {S.}~\bibnamefont {Hotchkiss}},\ }\bibfield  {title}
  {\bibinfo {title} {{WIMPs and stellar-mass primordial black holes are
  incompatible}},\ }\href {https://doi.org/10.1103/PhysRevD.100.023506}
  {\bibfield  {journal} {\bibinfo  {journal} {Phys. Rev. D}\ }\textbf {\bibinfo
  {volume} {100}},\ \bibinfo {pages} {023506} (\bibinfo {year} {2019})},\
  \Eprint {https://arxiv.org/abs/1901.08528} {arXiv:1901.08528 [astro-ph.CO]}
  \BibitemShut {NoStop}%
\bibitem [{\citenamefont {Boucenna}\ \emph {et~al.}(2018)\citenamefont
  {Boucenna}, \citenamefont {Kuhnel}, \citenamefont {Ohlsson},\ and\
  \citenamefont {Visinelli}}]{Boucenna:2017ghj}%
  \BibitemOpen
  \bibfield  {author} {\bibinfo {author} {\bibfnamefont {S.~M.}\ \bibnamefont
  {Boucenna}}, \bibinfo {author} {\bibfnamefont {F.}~\bibnamefont {Kuhnel}},
  \bibinfo {author} {\bibfnamefont {T.}~\bibnamefont {Ohlsson}},\ and\ \bibinfo
  {author} {\bibfnamefont {L.}~\bibnamefont {Visinelli}},\ }\bibfield  {title}
  {\bibinfo {title} {{Novel Constraints on Mixed Dark-Matter Scenarios of
  Primordial Black Holes and WIMPs}},\ }\href
  {https://doi.org/10.1088/1475-7516/2018/07/003} {\bibfield  {journal}
  {\bibinfo  {journal} {JCAP}\ }\textbf {\bibinfo {volume} {07}},\ \bibinfo
  {pages} {003}},\ \Eprint {https://arxiv.org/abs/1712.06383} {arXiv:1712.06383
  [hep-ph]} \BibitemShut {NoStop}%
\bibitem [{\citenamefont {Eroshenko}(2016)}]{Eroshenko:2016yve}%
  \BibitemOpen
  \bibfield  {author} {\bibinfo {author} {\bibfnamefont {Y.}~\bibnamefont
  {Eroshenko}},\ }\bibfield  {title} {\bibinfo {title} {{Dark matter density
  spikes around primordial black holes}},\ }\href
  {https://doi.org/10.1134/S1063773716060013} {\bibfield  {journal} {\bibinfo
  {journal} {Astron. Lett.}\ }\textbf {\bibinfo {volume} {42}},\ \bibinfo
  {pages} {347} (\bibinfo {year} {2016})},\ \Eprint
  {https://arxiv.org/abs/1607.00612} {arXiv:1607.00612 [astro-ph.HE]}
  \BibitemShut {NoStop}%
\bibitem [{\citenamefont {Carr}\ \emph {et~al.}(2021)\citenamefont {Carr},
  \citenamefont {Kuhnel},\ and\ \citenamefont {Visinelli}}]{Carr:2020mqm}%
  \BibitemOpen
  \bibfield  {author} {\bibinfo {author} {\bibfnamefont {B.}~\bibnamefont
  {Carr}}, \bibinfo {author} {\bibfnamefont {F.}~\bibnamefont {Kuhnel}},\ and\
  \bibinfo {author} {\bibfnamefont {L.}~\bibnamefont {Visinelli}},\ }\bibfield
  {title} {\bibinfo {title} {{Black holes and WIMPs: all or nothing or
  something else}},\ }\href {https://doi.org/10.1093/mnras/stab1930} {\bibfield
   {journal} {\bibinfo  {journal} {Mon. Not. Roy. Astron. Soc.}\ }\textbf
  {\bibinfo {volume} {506}},\ \bibinfo {pages} {3648} (\bibinfo {year}
  {2021})},\ \Eprint {https://arxiv.org/abs/2011.01930} {arXiv:2011.01930
  [astro-ph.CO]} \BibitemShut {NoStop}%
\bibitem [{\citenamefont {Cai}\ \emph {et~al.}(2020)\citenamefont {Cai},
  \citenamefont {Yang},\ and\ \citenamefont {Zhou}}]{Cai:2020fnq}%
  \BibitemOpen
  \bibfield  {author} {\bibinfo {author} {\bibfnamefont {R.-G.}\ \bibnamefont
  {Cai}}, \bibinfo {author} {\bibfnamefont {X.-Y.}\ \bibnamefont {Yang}},\ and\
  \bibinfo {author} {\bibfnamefont {Y.-F.}\ \bibnamefont {Zhou}},\ }\bibfield
  {title} {\bibinfo {title} {{Constraints on mixed dark matter model of
  particles and primordial black holes from the Galactic 511 keV line}},\
  }\href@noop {} {\  (\bibinfo {year} {2020})},\ \Eprint
  {https://arxiv.org/abs/2007.11804} {arXiv:2007.11804 [astro-ph.CO]}
  \BibitemShut {NoStop}%
\bibitem [{\citenamefont {Delos}\ \emph {et~al.}(2018)\citenamefont {Delos},
  \citenamefont {Erickcek}, \citenamefont {Bailey},\ and\ \citenamefont
  {Alvarez}}]{Delos:2018ueo}%
  \BibitemOpen
  \bibfield  {author} {\bibinfo {author} {\bibfnamefont {M.~S.}\ \bibnamefont
  {Delos}}, \bibinfo {author} {\bibfnamefont {A.~L.}\ \bibnamefont {Erickcek}},
  \bibinfo {author} {\bibfnamefont {A.~P.}\ \bibnamefont {Bailey}},\ and\
  \bibinfo {author} {\bibfnamefont {M.~A.}\ \bibnamefont {Alvarez}},\
  }\bibfield  {title} {\bibinfo {title} {{Density profiles of ultracompact
  minihalos: Implications for constraining the primordial power spectrum}},\
  }\href {https://doi.org/10.1103/PhysRevD.98.063527} {\bibfield  {journal}
  {\bibinfo  {journal} {Phys. Rev. D}\ }\textbf {\bibinfo {volume} {98}},\
  \bibinfo {pages} {063527} (\bibinfo {year} {2018})},\ \Eprint
  {https://arxiv.org/abs/1806.07389} {arXiv:1806.07389 [astro-ph.CO]}
  \BibitemShut {NoStop}%
\bibitem [{\citenamefont {Kohri}\ \emph {et~al.}(2014)\citenamefont {Kohri},
  \citenamefont {Nakama},\ and\ \citenamefont {Suyama}}]{Kohri:2014lza}%
  \BibitemOpen
  \bibfield  {author} {\bibinfo {author} {\bibfnamefont {K.}~\bibnamefont
  {Kohri}}, \bibinfo {author} {\bibfnamefont {T.}~\bibnamefont {Nakama}},\ and\
  \bibinfo {author} {\bibfnamefont {T.}~\bibnamefont {Suyama}},\ }\bibfield
  {title} {\bibinfo {title} {{Testing scenarios of primordial black holes being
  the seeds of supermassive black holes by ultracompact minihalos and CMB
  $\mu$-distortions}},\ }\href {https://doi.org/10.1103/PhysRevD.90.083514}
  {\bibfield  {journal} {\bibinfo  {journal} {Phys. Rev. D}\ }\textbf {\bibinfo
  {volume} {90}},\ \bibinfo {pages} {083514} (\bibinfo {year} {2014})},\
  \Eprint {https://arxiv.org/abs/1405.5999} {arXiv:1405.5999 [astro-ph.CO]}
  \BibitemShut {NoStop}%
\bibitem [{\citenamefont {Bertone}\ \emph {et~al.}(2019)\citenamefont
  {Bertone}, \citenamefont {Coogan}, \citenamefont {Gaggero}, \citenamefont
  {Kavanagh},\ and\ \citenamefont {Weniger}}]{Bertone:2019vsk}%
  \BibitemOpen
  \bibfield  {author} {\bibinfo {author} {\bibfnamefont {G.}~\bibnamefont
  {Bertone}}, \bibinfo {author} {\bibfnamefont {A.~M.}\ \bibnamefont {Coogan}},
  \bibinfo {author} {\bibfnamefont {D.}~\bibnamefont {Gaggero}}, \bibinfo
  {author} {\bibfnamefont {B.~J.}\ \bibnamefont {Kavanagh}},\ and\ \bibinfo
  {author} {\bibfnamefont {C.}~\bibnamefont {Weniger}},\ }\bibfield  {title}
  {\bibinfo {title} {{Primordial Black Holes as Silver Bullets for New Physics
  at the Weak Scale}},\ }\href {https://doi.org/10.1103/PhysRevD.100.123013}
  {\bibfield  {journal} {\bibinfo  {journal} {Phys. Rev. D}\ }\textbf {\bibinfo
  {volume} {100}},\ \bibinfo {pages} {123013} (\bibinfo {year} {2019})},\
  \Eprint {https://arxiv.org/abs/1905.01238} {arXiv:1905.01238 [hep-ph]}
  \BibitemShut {NoStop}%
\bibitem [{\citenamefont {Ando}\ and\ \citenamefont
  {Ishiwata}(2015)}]{Ando:2015qda}%
  \BibitemOpen
  \bibfield  {author} {\bibinfo {author} {\bibfnamefont {S.}~\bibnamefont
  {Ando}}\ and\ \bibinfo {author} {\bibfnamefont {K.}~\bibnamefont
  {Ishiwata}},\ }\bibfield  {title} {\bibinfo {title} {{Constraints on decaying
  dark matter from the extragalactic gamma-ray background}},\ }\href
  {https://doi.org/10.1088/1475-7516/2015/05/024} {\bibfield  {journal}
  {\bibinfo  {journal} {JCAP}\ }\textbf {\bibinfo {volume} {05}},\ \bibinfo
  {pages} {024}},\ \Eprint {https://arxiv.org/abs/1502.02007} {arXiv:1502.02007
  [astro-ph.CO]} \BibitemShut {NoStop}%
\bibitem [{\citenamefont {Hertzberg}\ \emph {et~al.}(2020)\citenamefont
  {Hertzberg}, \citenamefont {Nurmi}, \citenamefont {Schiappacasse},\ and\
  \citenamefont {Yanagida}}]{Hertzberg:2020kpm}%
  \BibitemOpen
  \bibfield  {author} {\bibinfo {author} {\bibfnamefont {M.~P.}\ \bibnamefont
  {Hertzberg}}, \bibinfo {author} {\bibfnamefont {S.}~\bibnamefont {Nurmi}},
  \bibinfo {author} {\bibfnamefont {E.~D.}\ \bibnamefont {Schiappacasse}},\
  and\ \bibinfo {author} {\bibfnamefont {T.~T.}\ \bibnamefont {Yanagida}},\
  }\bibfield  {title} {\bibinfo {title} {{Shining Primordial Black Holes}},\
  }\href@noop {} {\  (\bibinfo {year} {2020})},\ \Eprint
  {https://arxiv.org/abs/2011.05922} {arXiv:2011.05922 [hep-ph]} \BibitemShut
  {NoStop}%
\bibitem [{\citenamefont {Yang}(2020)}]{Yang:2020zcu}%
  \BibitemOpen
  \bibfield  {author} {\bibinfo {author} {\bibfnamefont {Y.}~\bibnamefont
  {Yang}},\ }\bibfield  {title} {\bibinfo {title} {{The abundance of primordial
  black holes from the global 21cm signal and extragalactic gamma-ray
  background}},\ }\href {https://doi.org/10.1140/epjp/s13360-020-00710-3}
  {\bibfield  {journal} {\bibinfo  {journal} {Eur. Phys. J. Plus}\ }\textbf
  {\bibinfo {volume} {135}},\ \bibinfo {pages} {690} (\bibinfo {year}
  {2020})},\ \Eprint {https://arxiv.org/abs/2008.11859} {arXiv:2008.11859
  [astro-ph.CO]} \BibitemShut {NoStop}%
\bibitem [{\citenamefont {Zhang}(2011)}]{Zhang:2010cj}%
  \BibitemOpen
  \bibfield  {author} {\bibinfo {author} {\bibfnamefont {D.}~\bibnamefont
  {Zhang}},\ }\bibfield  {title} {\bibinfo {title} {{Impact of Primordial
  Ultracompact Minihaloes on the Intergalactic Medium and First Structure
  Formation}},\ }\href {https://doi.org/10.1111/j.1365-2966.2011.19602.x}
  {\bibfield  {journal} {\bibinfo  {journal} {Mon. Not. Roy. Astron. Soc.}\
  }\textbf {\bibinfo {volume} {418}},\ \bibinfo {pages} {1850} (\bibinfo {year}
  {2011})},\ \Eprint {https://arxiv.org/abs/1011.1935} {arXiv:1011.1935
  [astro-ph.CO]} \BibitemShut {NoStop}%
\bibitem [{\citenamefont {Tashiro}\ and\ \citenamefont
  {Kadota}(2021)}]{Tashiro:2021xnj}%
  \BibitemOpen
  \bibfield  {author} {\bibinfo {author} {\bibfnamefont {H.}~\bibnamefont
  {Tashiro}}\ and\ \bibinfo {author} {\bibfnamefont {K.}~\bibnamefont
  {Kadota}},\ }\bibfield  {title} {\bibinfo {title} {{Constraining Mixed
  Dark-Matter Scenarios of WIMPs and Primordial Black Holes from CMB and 21-cm
  observations}},\ }\href {https://doi.org/10.1103/PhysRevD.103.123532}
  {\bibfield  {journal} {\bibinfo  {journal} {Phys. Rev. D}\ }\textbf {\bibinfo
  {volume} {103}},\ \bibinfo {pages} {123532} (\bibinfo {year} {2021})},\
  \Eprint {https://arxiv.org/abs/2104.09738} {arXiv:2104.09738 [astro-ph.CO]}
  \BibitemShut {NoStop}%
\bibitem [{\citenamefont {Kadota}\ and\ \citenamefont
  {Tashiro}(2022)}]{Kadota:2022cij}%
  \BibitemOpen
  \bibfield  {author} {\bibinfo {author} {\bibfnamefont {K.}~\bibnamefont
  {Kadota}}\ and\ \bibinfo {author} {\bibfnamefont {H.}~\bibnamefont
  {Tashiro}},\ }\bibfield  {title} {\bibinfo {title} {{Radio bounds on the
  mixed dark matter scenarios of primordial black holes and WIMPs}},\ }\href
  {https://doi.org/10.1088/1475-7516/2022/08/004} {\bibfield  {journal}
  {\bibinfo  {journal} {JCAP}\ }\textbf {\bibinfo {volume} {08}}\bibfield
  {number} {\bibinfo  {number} { (08)},\ \bibinfo {pages} {004}},\ }\Eprint
  {https://arxiv.org/abs/2204.13273} {arXiv:2204.13273 [hep-ph]} \BibitemShut
  {NoStop}%
\bibitem [{\citenamefont {Eda}\ \emph {et~al.}(2013)\citenamefont {Eda},
  \citenamefont {Itoh}, \citenamefont {Kuroyanagi},\ and\ \citenamefont
  {Silk}}]{Eda:2013gg}%
  \BibitemOpen
  \bibfield  {author} {\bibinfo {author} {\bibfnamefont {K.}~\bibnamefont
  {Eda}}, \bibinfo {author} {\bibfnamefont {Y.}~\bibnamefont {Itoh}}, \bibinfo
  {author} {\bibfnamefont {S.}~\bibnamefont {Kuroyanagi}},\ and\ \bibinfo
  {author} {\bibfnamefont {J.}~\bibnamefont {Silk}},\ }\bibfield  {title}
  {\bibinfo {title} {{New Probe of Dark-Matter Properties: Gravitational Waves
  from an Intermediate-Mass Black Hole Embedded in a Dark-Matter Minispike}},\
  }\href {https://doi.org/10.1103/PhysRevLett.110.221101} {\bibfield  {journal}
  {\bibinfo  {journal} {Phys. Rev. Lett.}\ }\textbf {\bibinfo {volume} {110}},\
  \bibinfo {pages} {221101} (\bibinfo {year} {2013})},\ \Eprint
  {https://arxiv.org/abs/1301.5971} {arXiv:1301.5971 [gr-qc]} \BibitemShut
  {NoStop}%
\bibitem [{\citenamefont {Eda}\ \emph {et~al.}(2015)\citenamefont {Eda},
  \citenamefont {Itoh}, \citenamefont {Kuroyanagi},\ and\ \citenamefont
  {Silk}}]{Eda:2014kra}%
  \BibitemOpen
  \bibfield  {author} {\bibinfo {author} {\bibfnamefont {K.}~\bibnamefont
  {Eda}}, \bibinfo {author} {\bibfnamefont {Y.}~\bibnamefont {Itoh}}, \bibinfo
  {author} {\bibfnamefont {S.}~\bibnamefont {Kuroyanagi}},\ and\ \bibinfo
  {author} {\bibfnamefont {J.}~\bibnamefont {Silk}},\ }\bibfield  {title}
  {\bibinfo {title} {{Gravitational waves as a probe of dark matter
  minispikes}},\ }\href {https://doi.org/10.1103/PhysRevD.91.044045} {\bibfield
   {journal} {\bibinfo  {journal} {Phys. Rev. D}\ }\textbf {\bibinfo {volume}
  {91}},\ \bibinfo {pages} {044045} (\bibinfo {year} {2015})},\ \Eprint
  {https://arxiv.org/abs/1408.3534} {arXiv:1408.3534 [gr-qc]} \BibitemShut
  {NoStop}%
\bibitem [{\citenamefont {Kadota}\ and\ \citenamefont
  {Silk}(2021)}]{Kadota:2020ahr}%
  \BibitemOpen
  \bibfield  {author} {\bibinfo {author} {\bibfnamefont {K.}~\bibnamefont
  {Kadota}}\ and\ \bibinfo {author} {\bibfnamefont {J.}~\bibnamefont {Silk}},\
  }\bibfield  {title} {\bibinfo {title} {{Boosting small-scale structure via
  primordial black holes and implications for sub-GeV dark matter
  annihilation}},\ }\href {https://doi.org/10.1103/PhysRevD.103.043530}
  {\bibfield  {journal} {\bibinfo  {journal} {Phys. Rev. D}\ }\textbf {\bibinfo
  {volume} {103}},\ \bibinfo {pages} {043530} (\bibinfo {year} {2021})},\
  \Eprint {https://arxiv.org/abs/2012.03698} {arXiv:2012.03698 [astro-ph.CO]}
  \BibitemShut {NoStop}%
\bibitem [{\citenamefont {Yue}\ and\ \citenamefont {Han}(2018)}]{Yue:2017iwc}%
  \BibitemOpen
  \bibfield  {author} {\bibinfo {author} {\bibfnamefont {X.-J.}\ \bibnamefont
  {Yue}}\ and\ \bibinfo {author} {\bibfnamefont {W.-B.}\ \bibnamefont {Han}},\
  }\bibfield  {title} {\bibinfo {title} {{Gravitational waves with dark matter
  minispikes: the combined effect}},\ }\href
  {https://doi.org/10.1103/PhysRevD.97.064003} {\bibfield  {journal} {\bibinfo
  {journal} {Phys. Rev. D}\ }\textbf {\bibinfo {volume} {97}},\ \bibinfo
  {pages} {064003} (\bibinfo {year} {2018})},\ \Eprint
  {https://arxiv.org/abs/1711.09706} {arXiv:1711.09706 [gr-qc]} \BibitemShut
  {NoStop}%
\bibitem [{\citenamefont {Macedo}\ \emph {et~al.}(2013)\citenamefont {Macedo},
  \citenamefont {Pani}, \citenamefont {Cardoso},\ and\ \citenamefont
  {Crispino}}]{Macedo:2013qea}%
  \BibitemOpen
  \bibfield  {author} {\bibinfo {author} {\bibfnamefont {C.~F.~B.}\
  \bibnamefont {Macedo}}, \bibinfo {author} {\bibfnamefont {P.}~\bibnamefont
  {Pani}}, \bibinfo {author} {\bibfnamefont {V.}~\bibnamefont {Cardoso}},\ and\
  \bibinfo {author} {\bibfnamefont {L.~C.~B.}\ \bibnamefont {Crispino}},\
  }\bibfield  {title} {\bibinfo {title} {{Into the lair: gravitational-wave
  signatures of dark matter}},\ }\href
  {https://doi.org/10.1088/0004-637X/774/1/48} {\bibfield  {journal} {\bibinfo
  {journal} {Astrophys. J.}\ }\textbf {\bibinfo {volume} {774}},\ \bibinfo
  {pages} {48} (\bibinfo {year} {2013})},\ \Eprint
  {https://arxiv.org/abs/1302.2646} {arXiv:1302.2646 [gr-qc]} \BibitemShut
  {NoStop}%
\bibitem [{\citenamefont {Barausse}\ \emph {et~al.}(2014)\citenamefont
  {Barausse}, \citenamefont {Cardoso},\ and\ \citenamefont
  {Pani}}]{Barausse:2014tra}%
  \BibitemOpen
  \bibfield  {author} {\bibinfo {author} {\bibfnamefont {E.}~\bibnamefont
  {Barausse}}, \bibinfo {author} {\bibfnamefont {V.}~\bibnamefont {Cardoso}},\
  and\ \bibinfo {author} {\bibfnamefont {P.}~\bibnamefont {Pani}},\ }\bibfield
  {title} {\bibinfo {title} {{Can environmental effects spoil precision
  gravitational-wave astrophysics?}},\ }\href
  {https://doi.org/10.1103/PhysRevD.89.104059} {\bibfield  {journal} {\bibinfo
  {journal} {Phys. Rev. D}\ }\textbf {\bibinfo {volume} {89}},\ \bibinfo
  {pages} {104059} (\bibinfo {year} {2014})},\ \Eprint
  {https://arxiv.org/abs/1404.7149} {arXiv:1404.7149 [gr-qc]} \BibitemShut
  {NoStop}%
\bibitem [{\citenamefont {Bertone}\ \emph {et~al.}(2020)\citenamefont {Bertone}
  \emph {et~al.}}]{Bertone:2019irm}%
  \BibitemOpen
  \bibfield  {author} {\bibinfo {author} {\bibfnamefont {G.}~\bibnamefont
  {Bertone}} \emph {et~al.},\ }\bibfield  {title} {\bibinfo {title}
  {{Gravitational wave probes of dark matter: challenges and opportunities}},\
  }\href {https://doi.org/10.21468/SciPostPhysCore.3.2.007} {\bibfield
  {journal} {\bibinfo  {journal} {SciPost Phys. Core}\ }\textbf {\bibinfo
  {volume} {3}},\ \bibinfo {pages} {007} (\bibinfo {year} {2020})},\ \Eprint
  {https://arxiv.org/abs/1907.10610} {arXiv:1907.10610 [astro-ph.CO]}
  \BibitemShut {NoStop}%
\bibitem [{\citenamefont {Cole}\ \emph {et~al.}(2022)\citenamefont {Cole},
  \citenamefont {Bertone}, \citenamefont {Coogan}, \citenamefont {Gaggero},
  \citenamefont {Karydas}, \citenamefont {Kavanagh}, \citenamefont {Spieksma},\
  and\ \citenamefont {Tomaselli}}]{Cole:2022fir}%
  \BibitemOpen
  \bibfield  {author} {\bibinfo {author} {\bibfnamefont {P.~S.}\ \bibnamefont
  {Cole}}, \bibinfo {author} {\bibfnamefont {G.}~\bibnamefont {Bertone}},
  \bibinfo {author} {\bibfnamefont {A.}~\bibnamefont {Coogan}}, \bibinfo
  {author} {\bibfnamefont {D.}~\bibnamefont {Gaggero}}, \bibinfo {author}
  {\bibfnamefont {T.}~\bibnamefont {Karydas}}, \bibinfo {author} {\bibfnamefont
  {B.~J.}\ \bibnamefont {Kavanagh}}, \bibinfo {author} {\bibfnamefont
  {T.~F.~M.}\ \bibnamefont {Spieksma}},\ and\ \bibinfo {author} {\bibfnamefont
  {G.~M.}\ \bibnamefont {Tomaselli}},\ }\bibfield  {title} {\bibinfo {title}
  {{Disks, spikes, and clouds: distinguishing environmental effects on BBH
  gravitational waveforms}},\ }\href@noop {} {\  (\bibinfo {year} {2022})},\
  \Eprint {https://arxiv.org/abs/2211.01362} {arXiv:2211.01362 [gr-qc]}
  \BibitemShut {NoStop}%
\bibitem [{\citenamefont {Kavanagh}\ \emph {et~al.}(2020)\citenamefont
  {Kavanagh}, \citenamefont {Nichols}, \citenamefont {Bertone},\ and\
  \citenamefont {Gaggero}}]{Kavanagh:2020cfn}%
  \BibitemOpen
  \bibfield  {author} {\bibinfo {author} {\bibfnamefont {B.~J.}\ \bibnamefont
  {Kavanagh}}, \bibinfo {author} {\bibfnamefont {D.~A.}\ \bibnamefont
  {Nichols}}, \bibinfo {author} {\bibfnamefont {G.}~\bibnamefont {Bertone}},\
  and\ \bibinfo {author} {\bibfnamefont {D.}~\bibnamefont {Gaggero}},\
  }\bibfield  {title} {\bibinfo {title} {{Detecting dark matter around black
  holes with gravitational waves: Effects of dark-matter dynamics on the
  gravitational waveform}},\ }\href
  {https://doi.org/10.1103/PhysRevD.102.083006} {\bibfield  {journal} {\bibinfo
   {journal} {Phys. Rev. D}\ }\textbf {\bibinfo {volume} {102}},\ \bibinfo
  {pages} {083006} (\bibinfo {year} {2020})},\ \Eprint
  {https://arxiv.org/abs/2002.12811} {arXiv:2002.12811 [gr-qc]} \BibitemShut
  {NoStop}%
\bibitem [{\citenamefont {Cardoso}\ and\ \citenamefont
  {Maselli}(2020)}]{Cardoso:2019rou}%
  \BibitemOpen
  \bibfield  {author} {\bibinfo {author} {\bibfnamefont {V.}~\bibnamefont
  {Cardoso}}\ and\ \bibinfo {author} {\bibfnamefont {A.}~\bibnamefont
  {Maselli}},\ }\bibfield  {title} {\bibinfo {title} {{Constraints on the
  astrophysical environment of binaries with gravitational-wave
  observations}},\ }\href {https://doi.org/10.1051/0004-6361/202037654}
  {\bibfield  {journal} {\bibinfo  {journal} {Astron. Astrophys.}\ }\textbf
  {\bibinfo {volume} {644}},\ \bibinfo {pages} {A147} (\bibinfo {year}
  {2020})},\ \Eprint {https://arxiv.org/abs/1909.05870} {arXiv:1909.05870
  [astro-ph.HE]} \BibitemShut {NoStop}%
\bibitem [{\citenamefont {Hannuksela}\ \emph {et~al.}(2020)\citenamefont
  {Hannuksela}, \citenamefont {Ng},\ and\ \citenamefont
  {Li}}]{Hannuksela:2019vip}%
  \BibitemOpen
  \bibfield  {author} {\bibinfo {author} {\bibfnamefont {O.~A.}\ \bibnamefont
  {Hannuksela}}, \bibinfo {author} {\bibfnamefont {K.~C.~Y.}\ \bibnamefont
  {Ng}},\ and\ \bibinfo {author} {\bibfnamefont {T.~G.~F.}\ \bibnamefont
  {Li}},\ }\bibfield  {title} {\bibinfo {title} {{Extreme dark matter tests
  with extreme mass ratio inspirals}},\ }\href
  {https://doi.org/10.1103/PhysRevD.102.103022} {\bibfield  {journal} {\bibinfo
   {journal} {Phys. Rev. D}\ }\textbf {\bibinfo {volume} {102}},\ \bibinfo
  {pages} {103022} (\bibinfo {year} {2020})},\ \Eprint
  {https://arxiv.org/abs/1906.11845} {arXiv:1906.11845 [astro-ph.CO]}
  \BibitemShut {NoStop}%
\bibitem [{\citenamefont {Coogan}\ \emph {et~al.}(2021)\citenamefont {Coogan},
  \citenamefont {Bertone}, \citenamefont {Gaggero}, \citenamefont {Kavanagh},\
  and\ \citenamefont {Nichols}}]{Coogan:2021uqv}%
  \BibitemOpen
  \bibfield  {author} {\bibinfo {author} {\bibfnamefont {A.}~\bibnamefont
  {Coogan}}, \bibinfo {author} {\bibfnamefont {G.}~\bibnamefont {Bertone}},
  \bibinfo {author} {\bibfnamefont {D.}~\bibnamefont {Gaggero}}, \bibinfo
  {author} {\bibfnamefont {B.~J.}\ \bibnamefont {Kavanagh}},\ and\ \bibinfo
  {author} {\bibfnamefont {D.~A.}\ \bibnamefont {Nichols}},\ }\bibfield
  {title} {\bibinfo {title} {{Measuring the dark matter environments of black
  hole binaries with gravitational waves}},\ }\href@noop {} {\  (\bibinfo
  {year} {2021})},\ \Eprint {https://arxiv.org/abs/2108.04154}
  {arXiv:2108.04154 [gr-qc]} \BibitemShut {NoStop}%
\bibitem [{\citenamefont {Kim}\ \emph {et~al.}(2023)\citenamefont {Kim},
  \citenamefont {Lenoci}, \citenamefont {Stomberg},\ and\ \citenamefont
  {Xue}}]{Kim:2022mdj}%
  \BibitemOpen
  \bibfield  {author} {\bibinfo {author} {\bibfnamefont {H.}~\bibnamefont
  {Kim}}, \bibinfo {author} {\bibfnamefont {A.}~\bibnamefont {Lenoci}},
  \bibinfo {author} {\bibfnamefont {I.}~\bibnamefont {Stomberg}},\ and\
  \bibinfo {author} {\bibfnamefont {X.}~\bibnamefont {Xue}},\ }\bibfield
  {title} {\bibinfo {title} {{Adiabatically compressed wave dark matter halo
  and intermediate-mass-ratio inspirals}},\ }\href
  {https://doi.org/10.1103/PhysRevD.107.083005} {\bibfield  {journal} {\bibinfo
   {journal} {Phys. Rev. D}\ }\textbf {\bibinfo {volume} {107}},\ \bibinfo
  {pages} {083005} (\bibinfo {year} {2023})},\ \Eprint
  {https://arxiv.org/abs/2212.07528} {arXiv:2212.07528 [astro-ph.GA]}
  \BibitemShut {NoStop}%
\bibitem [{\citenamefont {Kadota}\ \emph {et~al.}(2023)\citenamefont {Kadota},
  \citenamefont {Kim}, \citenamefont {Ko},\ and\ \citenamefont
  {Yang}}]{Kadota:2023wlm}%
  \BibitemOpen
  \bibfield  {author} {\bibinfo {author} {\bibfnamefont {K.}~\bibnamefont
  {Kadota}}, \bibinfo {author} {\bibfnamefont {J.~H.}\ \bibnamefont {Kim}},
  \bibinfo {author} {\bibfnamefont {P.}~\bibnamefont {Ko}},\ and\ \bibinfo
  {author} {\bibfnamefont {X.-Y.}\ \bibnamefont {Yang}},\ }\bibfield  {title}
  {\bibinfo {title} {{Gravitational Wave Probes on Self-Interacting Dark Matter
  Surrounding an Intermediate Mass Black Hole}},\ }\href@noop {} {\  (\bibinfo
  {year} {2023})},\ \Eprint {https://arxiv.org/abs/2306.10828}
  {arXiv:2306.10828 [hep-ph]} \BibitemShut {NoStop}%
\bibitem [{ska()}]{skawebpage}%
  \BibitemOpen
  \href {https://www.skao.int/} {\bibinfo {title} {ska webpage:}},\ \bibinfo
  {note} {\url{https://www.skao.int/}}\BibitemShut {NoStop}%
\bibitem [{\citenamefont {{Braun}}\ \emph {et~al.}(2019)\citenamefont
  {{Braun}}, \citenamefont {{Bonaldi}}, \citenamefont {{Bourke}}, \citenamefont
  {{Keane}},\ and\ \citenamefont {{Wagg}}}]{2019arXiv191212699B}%
  \BibitemOpen
  \bibfield  {author} {\bibinfo {author} {\bibfnamefont {R.}~\bibnamefont
  {{Braun}}}, \bibinfo {author} {\bibfnamefont {A.}~\bibnamefont {{Bonaldi}}},
  \bibinfo {author} {\bibfnamefont {T.}~\bibnamefont {{Bourke}}}, \bibinfo
  {author} {\bibfnamefont {E.}~\bibnamefont {{Keane}}},\ and\ \bibinfo {author}
  {\bibfnamefont {J.}~\bibnamefont {{Wagg}}},\ }\bibfield  {title} {\bibinfo
  {title} {{Anticipated Performance of the Square Kilometre Array -- Phase 1
  (SKA1)}},\ }\href@noop {} {\bibfield  {journal} {\bibinfo  {journal} {arXiv
  e-prints}\ ,\ \bibinfo {eid} {arXiv:1912.12699}} (\bibinfo {year}
  {2019})}\BibitemShut {NoStop}%
\bibitem [{\citenamefont {Cembranos}\ \emph {et~al.}(2020)\citenamefont
  {Cembranos}, \citenamefont {De~La Cruz-Dombriz}, \citenamefont {Gammaldi},\
  and\ \citenamefont {M\'endez-Isla}}]{Cembranos:2019noa}%
  \BibitemOpen
  \bibfield  {author} {\bibinfo {author} {\bibfnamefont {J.~A.~R.}\
  \bibnamefont {Cembranos}}, \bibinfo {author} {\bibfnamefont {A.}~\bibnamefont
  {De~La Cruz-Dombriz}}, \bibinfo {author} {\bibfnamefont {V.}~\bibnamefont
  {Gammaldi}},\ and\ \bibinfo {author} {\bibfnamefont {M.}~\bibnamefont
  {M\'endez-Isla}},\ }\bibfield  {title} {\bibinfo {title} {{SKA-Phase 1
  sensitivity to synchrotron radio emission from multi-TeV Dark Matter
  candidates}},\ }\href {https://doi.org/10.1016/j.dark.2019.100448} {\bibfield
   {journal} {\bibinfo  {journal} {Phys. Dark Univ.}\ }\textbf {\bibinfo
  {volume} {27}},\ \bibinfo {pages} {100448} (\bibinfo {year} {2020})},\
  \Eprint {https://arxiv.org/abs/1905.11154} {arXiv:1905.11154 [hep-ph]}
  \BibitemShut {NoStop}%
\bibitem [{\citenamefont {Colafrancesco}\ \emph {et~al.}(2015)\citenamefont
  {Colafrancesco}, \citenamefont {Regis}, \citenamefont {Marchegiani},
  \citenamefont {Beck}, \citenamefont {Beck}, \citenamefont {Zechlin},
  \citenamefont {Lobanov},\ and\ \citenamefont
  {Horns}}]{Colafrancesco:2015ola}%
  \BibitemOpen
  \bibfield  {author} {\bibinfo {author} {\bibfnamefont {S.}~\bibnamefont
  {Colafrancesco}}, \bibinfo {author} {\bibfnamefont {M.}~\bibnamefont
  {Regis}}, \bibinfo {author} {\bibfnamefont {P.}~\bibnamefont {Marchegiani}},
  \bibinfo {author} {\bibfnamefont {G.}~\bibnamefont {Beck}}, \bibinfo {author}
  {\bibfnamefont {R.}~\bibnamefont {Beck}}, \bibinfo {author} {\bibfnamefont
  {H.}~\bibnamefont {Zechlin}}, \bibinfo {author} {\bibfnamefont
  {A.}~\bibnamefont {Lobanov}},\ and\ \bibinfo {author} {\bibfnamefont
  {D.}~\bibnamefont {Horns}},\ }\bibfield  {title} {\bibinfo {title} {{Probing
  the nature of Dark Matter with the SKA}},\ }\href
  {https://doi.org/10.22323/1.215.0100} {\bibfield  {journal} {\bibinfo
  {journal} {PoS}\ }\textbf {\bibinfo {volume} {AASKA14}},\ \bibinfo {pages}
  {100} (\bibinfo {year} {2015})},\ \Eprint {https://arxiv.org/abs/1502.03738}
  {arXiv:1502.03738 [astro-ph.HE]} \BibitemShut {NoStop}%
\bibitem [{\citenamefont {Chen}\ \emph {et~al.}(2021)\citenamefont {Chen},
  \citenamefont {Tsai},\ and\ \citenamefont {Yuan}}]{Chen:2021rea}%
  \BibitemOpen
  \bibfield  {author} {\bibinfo {author} {\bibfnamefont {Z.}~\bibnamefont
  {Chen}}, \bibinfo {author} {\bibfnamefont {Y.-L.~S.}\ \bibnamefont {Tsai}},\
  and\ \bibinfo {author} {\bibfnamefont {Q.}~\bibnamefont {Yuan}},\ }\bibfield
  {title} {\bibinfo {title} {{Sensitivity of SKA to dark matter induced radio
  emission}},\ }\href {https://doi.org/10.1088/1475-7516/2021/09/025}
  {\bibfield  {journal} {\bibinfo  {journal} {JCAP}\ }\textbf {\bibinfo
  {volume} {09}},\ \bibinfo {pages} {025}},\ \Eprint
  {https://arxiv.org/abs/2105.00776} {arXiv:2105.00776 [astro-ph.HE]}
  \BibitemShut {NoStop}%
\bibitem [{\citenamefont {Wang}\ \emph {et~al.}(2023)\citenamefont {Wang},
  \citenamefont {Chen}, \citenamefont {Zu}, \citenamefont {Gong}, \citenamefont
  {Feng},\ and\ \citenamefont {Fan}}]{Wang:2023sxr}%
  \BibitemOpen
  \bibfield  {author} {\bibinfo {author} {\bibfnamefont {G.-S.}\ \bibnamefont
  {Wang}}, \bibinfo {author} {\bibfnamefont {Z.-F.}\ \bibnamefont {Chen}},
  \bibinfo {author} {\bibfnamefont {L.}~\bibnamefont {Zu}}, \bibinfo {author}
  {\bibfnamefont {H.}~\bibnamefont {Gong}}, \bibinfo {author} {\bibfnamefont
  {L.}~\bibnamefont {Feng}},\ and\ \bibinfo {author} {\bibfnamefont {Y.-Z.}\
  \bibnamefont {Fan}},\ }\bibfield  {title} {\bibinfo {title} {{SKA sensitivity
  for possible radio emission from dark matter in Omega Centauri}},\
  }\href@noop {} {\  (\bibinfo {year} {2023})},\ \Eprint
  {https://arxiv.org/abs/2303.14117} {arXiv:2303.14117 [astro-ph.HE]}
  \BibitemShut {NoStop}%
\bibitem [{\citenamefont {Ghosh}\ \emph {et~al.}(2020)\citenamefont {Ghosh},
  \citenamefont {Kar},\ and\ \citenamefont {Mukhopadhyaya}}]{Ghosh:2020ipv}%
  \BibitemOpen
  \bibfield  {author} {\bibinfo {author} {\bibfnamefont {A.}~\bibnamefont
  {Ghosh}}, \bibinfo {author} {\bibfnamefont {A.}~\bibnamefont {Kar}},\ and\
  \bibinfo {author} {\bibfnamefont {B.}~\bibnamefont {Mukhopadhyaya}},\
  }\bibfield  {title} {\bibinfo {title} {{Search for decaying heavy dark matter
  in an effective interaction framework: a comparison of $\gamma$-ray and radio
  observations}},\ }\href {https://doi.org/10.1088/1475-7516/2020/09/003}
  {\bibfield  {journal} {\bibinfo  {journal} {JCAP}\ }\textbf {\bibinfo
  {volume} {09}},\ \bibinfo {pages} {003}},\ \Eprint
  {https://arxiv.org/abs/2001.08235} {arXiv:2001.08235 [hep-ph]} \BibitemShut
  {NoStop}%
\bibitem [{\citenamefont {Colafrancesco}\ \emph {et~al.}(2006)\citenamefont
  {Colafrancesco}, \citenamefont {Profumo},\ and\ \citenamefont
  {Ullio}}]{Colafrancesco:2005ji}%
  \BibitemOpen
  \bibfield  {author} {\bibinfo {author} {\bibfnamefont {S.}~\bibnamefont
  {Colafrancesco}}, \bibinfo {author} {\bibfnamefont {S.}~\bibnamefont
  {Profumo}},\ and\ \bibinfo {author} {\bibfnamefont {P.}~\bibnamefont
  {Ullio}},\ }\bibfield  {title} {\bibinfo {title} {{Multi-frequency analysis
  of neutralino dark matter annihilations in the Coma cluster}},\ }\href
  {https://doi.org/10.1051/0004-6361:20053887} {\bibfield  {journal} {\bibinfo
  {journal} {Astron. Astrophys.}\ }\textbf {\bibinfo {volume} {455}},\ \bibinfo
  {pages} {21} (\bibinfo {year} {2006})},\ \Eprint
  {https://arxiv.org/abs/astro-ph/0507575} {arXiv:astro-ph/0507575}
  \BibitemShut {NoStop}%
\bibitem [{\citenamefont {McDaniel}\ \emph {et~al.}(2017)\citenamefont
  {McDaniel}, \citenamefont {Jeltema}, \citenamefont {Profumo},\ and\
  \citenamefont {Storm}}]{McDaniel:2017ppt}%
  \BibitemOpen
  \bibfield  {author} {\bibinfo {author} {\bibfnamefont {A.}~\bibnamefont
  {McDaniel}}, \bibinfo {author} {\bibfnamefont {T.}~\bibnamefont {Jeltema}},
  \bibinfo {author} {\bibfnamefont {S.}~\bibnamefont {Profumo}},\ and\ \bibinfo
  {author} {\bibfnamefont {E.}~\bibnamefont {Storm}},\ }\bibfield  {title}
  {\bibinfo {title} {{Multiwavelength Analysis of Dark Matter Annihilation and
  RX-DMFIT}},\ }\href {https://doi.org/10.1088/1475-7516/2017/09/027}
  {\bibfield  {journal} {\bibinfo  {journal} {JCAP}\ }\textbf {\bibinfo
  {volume} {09}},\ \bibinfo {pages} {027}},\ \Eprint
  {https://arxiv.org/abs/1705.09384} {arXiv:1705.09384 [astro-ph.HE]}
  \BibitemShut {NoStop}%
\bibitem [{\citenamefont {Dutta}\ \emph {et~al.}(2021)\citenamefont {Dutta},
  \citenamefont {Kar},\ and\ \citenamefont {Strigari}}]{Dutta:2020lqc}%
  \BibitemOpen
  \bibfield  {author} {\bibinfo {author} {\bibfnamefont {B.}~\bibnamefont
  {Dutta}}, \bibinfo {author} {\bibfnamefont {A.}~\bibnamefont {Kar}},\ and\
  \bibinfo {author} {\bibfnamefont {L.~E.}\ \bibnamefont {Strigari}},\
  }\bibfield  {title} {\bibinfo {title} {{Constraints on MeV dark matter and
  primordial black holes: Inverse Compton signals at the SKA}},\ }\href
  {https://doi.org/10.1088/1475-7516/2021/03/011} {\bibfield  {journal}
  {\bibinfo  {journal} {JCAP}\ }\textbf {\bibinfo {volume} {03}},\ \bibinfo
  {pages} {011}},\ \Eprint {https://arxiv.org/abs/2010.05977} {arXiv:2010.05977
  [astro-ph.HE]} \BibitemShut {NoStop}%
\bibitem [{\citenamefont {Ferrario}\ \emph {et~al.}(2015)\citenamefont
  {Ferrario}, \citenamefont {de~Martino},\ and\ \citenamefont
  {Gaensicke}}]{Ferrario:2015oda}%
  \BibitemOpen
  \bibfield  {author} {\bibinfo {author} {\bibfnamefont {L.}~\bibnamefont
  {Ferrario}}, \bibinfo {author} {\bibfnamefont {D.}~\bibnamefont
  {de~Martino}},\ and\ \bibinfo {author} {\bibfnamefont {B.}~\bibnamefont
  {Gaensicke}},\ }\bibfield  {title} {\bibinfo {title} {{Magnetic White
  Dwarfs}},\ }\href {https://doi.org/10.1007/s11214-015-0152-0} {\bibfield
  {journal} {\bibinfo  {journal} {Space Sci. Rev.}\ }\textbf {\bibinfo {volume}
  {191}},\ \bibinfo {pages} {111} (\bibinfo {year} {2015})},\ \Eprint
  {https://arxiv.org/abs/1504.08072} {arXiv:1504.08072 [astro-ph.SR]}
  \BibitemShut {NoStop}%
\bibitem [{\citenamefont {{Ferrario}}\ \emph {et~al.}(2020)\citenamefont
  {{Ferrario}}, \citenamefont {{Wickramasinghe}},\ and\ \citenamefont
  {{Kawka}}}]{2020AdSpR..66.1025F}%
  \BibitemOpen
  \bibfield  {author} {\bibinfo {author} {\bibfnamefont {L.}~\bibnamefont
  {{Ferrario}}}, \bibinfo {author} {\bibfnamefont {D.}~\bibnamefont
  {{Wickramasinghe}}},\ and\ \bibinfo {author} {\bibfnamefont {A.}~\bibnamefont
  {{Kawka}}},\ }\bibfield  {title} {\bibinfo {title} {{Magnetic fields in
  isolated and interacting white dwarfs}},\ }\href
  {https://doi.org/10.1016/j.asr.2019.11.012} {\bibfield  {journal} {\bibinfo
  {journal} {Advances in Space Research}\ }\textbf {\bibinfo {volume} {66}},\
  \bibinfo {pages} {1025} (\bibinfo {year} {2020})},\ \Eprint
  {https://arxiv.org/abs/2001.10147} {arXiv:2001.10147 [astro-ph.SR]}
  \BibitemShut {NoStop}%
\bibitem [{\citenamefont {{Bagnulo}}\ and\ \citenamefont
  {{Landstreet}}(2022)}]{2022ApJ...935L..12B}%
  \BibitemOpen
  \bibfield  {author} {\bibinfo {author} {\bibfnamefont {S.}~\bibnamefont
  {{Bagnulo}}}\ and\ \bibinfo {author} {\bibfnamefont {J.~D.}\ \bibnamefont
  {{Landstreet}}},\ }\bibfield  {title} {\bibinfo {title} {{Multiple Channels
  for the Onset of Magnetism in Isolated White Dwarfs}},\ }\href
  {https://doi.org/10.3847/2041-8213/ac84d3} {\bibfield  {journal} {\bibinfo
  {journal} {\apjl}\ }\textbf {\bibinfo {volume} {935}},\ \bibinfo {eid} {L12}
  (\bibinfo {year} {2022})},\ \Eprint {https://arxiv.org/abs/2208.02655}
  {arXiv:2208.02655 [astro-ph.SR]} \BibitemShut {NoStop}%
\bibitem [{\citenamefont {Boudaud}\ \emph {et~al.}(2021)\citenamefont
  {Boudaud}, \citenamefont {Lacroix}, \citenamefont {Stref}, \citenamefont
  {Lavalle},\ and\ \citenamefont {Salati}}]{Boudaud:2021irr}%
  \BibitemOpen
  \bibfield  {author} {\bibinfo {author} {\bibfnamefont {M.}~\bibnamefont
  {Boudaud}}, \bibinfo {author} {\bibfnamefont {T.}~\bibnamefont {Lacroix}},
  \bibinfo {author} {\bibfnamefont {M.}~\bibnamefont {Stref}}, \bibinfo
  {author} {\bibfnamefont {J.}~\bibnamefont {Lavalle}},\ and\ \bibinfo {author}
  {\bibfnamefont {P.}~\bibnamefont {Salati}},\ }\bibfield  {title} {\bibinfo
  {title} {{In-depth analysis of the clustering of dark matter particles around
  primordial black holes. Part~I. Density profiles}},\ }\href
  {https://doi.org/10.1088/1475-7516/2021/08/053} {\bibfield  {journal}
  {\bibinfo  {journal} {JCAP}\ }\textbf {\bibinfo {volume} {08}},\ \bibinfo
  {pages} {053}},\ \Eprint {https://arxiv.org/abs/2106.07480} {arXiv:2106.07480
  [astro-ph.CO]} \BibitemShut {NoStop}%
\bibitem [{\citenamefont {Chanda}\ \emph {et~al.}(2022)\citenamefont {Chanda},
  \citenamefont {Scholtz},\ and\ \citenamefont {Unwin}}]{Chanda:2022hls}%
  \BibitemOpen
  \bibfield  {author} {\bibinfo {author} {\bibfnamefont {P.}~\bibnamefont
  {Chanda}}, \bibinfo {author} {\bibfnamefont {J.}~\bibnamefont {Scholtz}},\
  and\ \bibinfo {author} {\bibfnamefont {J.}~\bibnamefont {Unwin}},\ }\bibfield
   {title} {\bibinfo {title} {{Improved Constraints on Dark Matter
  Annihilations Around Primordial Black Holes}},\ }\href@noop {} {\  (\bibinfo
  {year} {2022})},\ \Eprint {https://arxiv.org/abs/2209.07541}
  {arXiv:2209.07541 [hep-ph]} \BibitemShut {NoStop}%
\bibitem [{\citenamefont {Loeb}\ and\ \citenamefont
  {Zaldarriaga}(2005)}]{Loeb:2005pm}%
  \BibitemOpen
  \bibfield  {author} {\bibinfo {author} {\bibfnamefont {A.}~\bibnamefont
  {Loeb}}\ and\ \bibinfo {author} {\bibfnamefont {M.}~\bibnamefont
  {Zaldarriaga}},\ }\bibfield  {title} {\bibinfo {title} {{The Small-scale
  power spectrum of cold dark matter}},\ }\href
  {https://doi.org/10.1103/PhysRevD.71.103520} {\bibfield  {journal} {\bibinfo
  {journal} {Phys. Rev. D}\ }\textbf {\bibinfo {volume} {71}},\ \bibinfo
  {pages} {103520} (\bibinfo {year} {2005})},\ \Eprint
  {https://arxiv.org/abs/astro-ph/0504112} {arXiv:astro-ph/0504112}
  \BibitemShut {NoStop}%
\bibitem [{\citenamefont {Bertschinger}(2006)}]{Bertschinger:2006nq}%
  \BibitemOpen
  \bibfield  {author} {\bibinfo {author} {\bibfnamefont {E.}~\bibnamefont
  {Bertschinger}},\ }\bibfield  {title} {\bibinfo {title} {{The Effects of Cold
  Dark Matter Decoupling and Pair Annihilation on Cosmological
  Perturbations}},\ }\href {https://doi.org/10.1103/PhysRevD.74.063509}
  {\bibfield  {journal} {\bibinfo  {journal} {Phys. Rev. D}\ }\textbf {\bibinfo
  {volume} {74}},\ \bibinfo {pages} {063509} (\bibinfo {year} {2006})},\
  \Eprint {https://arxiv.org/abs/astro-ph/0607319} {arXiv:astro-ph/0607319}
  \BibitemShut {NoStop}%
\bibitem [{\citenamefont {Gondolo}\ \emph {et~al.}(2012)\citenamefont
  {Gondolo}, \citenamefont {Hisano},\ and\ \citenamefont
  {Kadota}}]{Gondolo:2012vh}%
  \BibitemOpen
  \bibfield  {author} {\bibinfo {author} {\bibfnamefont {P.}~\bibnamefont
  {Gondolo}}, \bibinfo {author} {\bibfnamefont {J.}~\bibnamefont {Hisano}},\
  and\ \bibinfo {author} {\bibfnamefont {K.}~\bibnamefont {Kadota}},\
  }\bibfield  {title} {\bibinfo {title} {{The Effect of quark interactions on
  dark matter kinetic decoupling and the mass of the smallest dark halos}},\
  }\href {https://doi.org/10.1103/PhysRevD.86.083523} {\bibfield  {journal}
  {\bibinfo  {journal} {Phys. Rev. D}\ }\textbf {\bibinfo {volume} {86}},\
  \bibinfo {pages} {083523} (\bibinfo {year} {2012})},\ \Eprint
  {https://arxiv.org/abs/1205.1914} {arXiv:1205.1914 [hep-ph]} \BibitemShut
  {NoStop}%
\bibitem [{\citenamefont {Profumo}\ \emph {et~al.}(2006)\citenamefont
  {Profumo}, \citenamefont {Sigurdson},\ and\ \citenamefont
  {Kamionkowski}}]{Profumo:2006bv}%
  \BibitemOpen
  \bibfield  {author} {\bibinfo {author} {\bibfnamefont {S.}~\bibnamefont
  {Profumo}}, \bibinfo {author} {\bibfnamefont {K.}~\bibnamefont {Sigurdson}},\
  and\ \bibinfo {author} {\bibfnamefont {M.}~\bibnamefont {Kamionkowski}},\
  }\bibfield  {title} {\bibinfo {title} {{What mass are the smallest
  protohalos?}},\ }\href {https://doi.org/10.1103/PhysRevLett.97.031301}
  {\bibfield  {journal} {\bibinfo  {journal} {Phys. Rev. Lett.}\ }\textbf
  {\bibinfo {volume} {97}},\ \bibinfo {pages} {031301} (\bibinfo {year}
  {2006})},\ \Eprint {https://arxiv.org/abs/astro-ph/0603373}
  {arXiv:astro-ph/0603373} \BibitemShut {NoStop}%
\bibitem [{\citenamefont {Gondolo}\ and\ \citenamefont
  {Kadota}(2016)}]{Gondolo:2016mrz}%
  \BibitemOpen
  \bibfield  {author} {\bibinfo {author} {\bibfnamefont {P.}~\bibnamefont
  {Gondolo}}\ and\ \bibinfo {author} {\bibfnamefont {K.}~\bibnamefont
  {Kadota}},\ }\bibfield  {title} {\bibinfo {title} {{Late Kinetic Decoupling
  of Light Magnetic Dipole Dark Matter}},\ }\href
  {https://doi.org/10.1088/1475-7516/2016/06/012} {\bibfield  {journal}
  {\bibinfo  {journal} {JCAP}\ }\textbf {\bibinfo {volume} {06}},\ \bibinfo
  {pages} {012}},\ \Eprint {https://arxiv.org/abs/1603.05783} {arXiv:1603.05783
  [hep-ph]} \BibitemShut {NoStop}%
\bibitem [{\citenamefont {Green}\ \emph {et~al.}(2004)\citenamefont {Green},
  \citenamefont {Hofmann},\ and\ \citenamefont {Schwarz}}]{Green:2003un}%
  \BibitemOpen
  \bibfield  {author} {\bibinfo {author} {\bibfnamefont {A.~M.}\ \bibnamefont
  {Green}}, \bibinfo {author} {\bibfnamefont {S.}~\bibnamefont {Hofmann}},\
  and\ \bibinfo {author} {\bibfnamefont {D.~J.}\ \bibnamefont {Schwarz}},\
  }\bibfield  {title} {\bibinfo {title} {{The power spectrum of SUSY - CDM on
  sub-galactic scales}},\ }\href
  {https://doi.org/10.1111/j.1365-2966.2004.08232.x} {\bibfield  {journal}
  {\bibinfo  {journal} {Mon. Not. Roy. Astron. Soc.}\ }\textbf {\bibinfo
  {volume} {353}},\ \bibinfo {pages} {L23} (\bibinfo {year} {2004})},\ \Eprint
  {https://arxiv.org/abs/astro-ph/0309621} {arXiv:astro-ph/0309621}
  \BibitemShut {NoStop}%
\bibitem [{\citenamefont {Green}\ \emph {et~al.}(2005)\citenamefont {Green},
  \citenamefont {Hofmann},\ and\ \citenamefont {Schwarz}}]{Green:2005fa}%
  \BibitemOpen
  \bibfield  {author} {\bibinfo {author} {\bibfnamefont {A.~M.}\ \bibnamefont
  {Green}}, \bibinfo {author} {\bibfnamefont {S.}~\bibnamefont {Hofmann}},\
  and\ \bibinfo {author} {\bibfnamefont {D.~J.}\ \bibnamefont {Schwarz}},\
  }\bibfield  {title} {\bibinfo {title} {{The First wimpy halos}},\ }\href
  {https://doi.org/10.1088/1475-7516/2005/08/003} {\bibfield  {journal}
  {\bibinfo  {journal} {JCAP}\ }\textbf {\bibinfo {volume} {08}},\ \bibinfo
  {pages} {003}},\ \Eprint {https://arxiv.org/abs/astro-ph/0503387}
  {arXiv:astro-ph/0503387} \BibitemShut {NoStop}%
\bibitem [{\citenamefont {Bringmann}\ and\ \citenamefont
  {Hofmann}(2007)}]{Bringmann:2006mu}%
  \BibitemOpen
  \bibfield  {author} {\bibinfo {author} {\bibfnamefont {T.}~\bibnamefont
  {Bringmann}}\ and\ \bibinfo {author} {\bibfnamefont {S.}~\bibnamefont
  {Hofmann}},\ }\bibfield  {title} {\bibinfo {title} {{Thermal decoupling of
  WIMPs from first principles}},\ }\href
  {https://doi.org/10.1088/1475-7516/2007/04/016} {\bibfield  {journal}
  {\bibinfo  {journal} {JCAP}\ }\textbf {\bibinfo {volume} {04}},\ \bibinfo
  {pages} {016}},\ \bibinfo {note} {[Erratum: JCAP 03, E02 (2016)]},\ \Eprint
  {https://arxiv.org/abs/hep-ph/0612238} {arXiv:hep-ph/0612238} \BibitemShut
  {NoStop}%
\bibitem [{\citenamefont {Vasiliev}(2007)}]{Vasiliev:2007vh}%
  \BibitemOpen
  \bibfield  {author} {\bibinfo {author} {\bibfnamefont {E.}~\bibnamefont
  {Vasiliev}},\ }\bibfield  {title} {\bibinfo {title} {{Dark matter
  annihilation near a black hole: Plateau vs. weak cusp}},\ }\href
  {https://doi.org/10.1103/PhysRevD.76.103532} {\bibfield  {journal} {\bibinfo
  {journal} {Phys. Rev. D}\ }\textbf {\bibinfo {volume} {76}},\ \bibinfo
  {pages} {103532} (\bibinfo {year} {2007})},\ \Eprint
  {https://arxiv.org/abs/0707.3334} {arXiv:0707.3334 [astro-ph]} \BibitemShut
  {NoStop}%
\bibitem [{\citenamefont {Sadeghian}\ \emph {et~al.}(2013)\citenamefont
  {Sadeghian}, \citenamefont {Ferrer},\ and\ \citenamefont
  {Will}}]{Sadeghian:2013laa}%
  \BibitemOpen
  \bibfield  {author} {\bibinfo {author} {\bibfnamefont {L.}~\bibnamefont
  {Sadeghian}}, \bibinfo {author} {\bibfnamefont {F.}~\bibnamefont {Ferrer}},\
  and\ \bibinfo {author} {\bibfnamefont {C.~M.}\ \bibnamefont {Will}},\
  }\bibfield  {title} {\bibinfo {title} {{Dark matter distributions around
  massive black holes: A general relativistic analysis}},\ }\href
  {https://doi.org/10.1103/PhysRevD.88.063522} {\bibfield  {journal} {\bibinfo
  {journal} {Phys. Rev. D}\ }\textbf {\bibinfo {volume} {88}},\ \bibinfo
  {pages} {063522} (\bibinfo {year} {2013})},\ \Eprint
  {https://arxiv.org/abs/1305.2619} {arXiv:1305.2619 [astro-ph.GA]}
  \BibitemShut {NoStop}%
\bibitem [{\citenamefont {Serpico}\ \emph {et~al.}(2020)\citenamefont
  {Serpico}, \citenamefont {Poulin}, \citenamefont {Inman},\ and\ \citenamefont
  {Kohri}}]{Serpico:2020ehh}%
  \BibitemOpen
  \bibfield  {author} {\bibinfo {author} {\bibfnamefont {P.~D.}\ \bibnamefont
  {Serpico}}, \bibinfo {author} {\bibfnamefont {V.}~\bibnamefont {Poulin}},
  \bibinfo {author} {\bibfnamefont {D.}~\bibnamefont {Inman}},\ and\ \bibinfo
  {author} {\bibfnamefont {K.}~\bibnamefont {Kohri}},\ }\bibfield  {title}
  {\bibinfo {title} {{Cosmic microwave background bounds on primordial black
  holes including dark matter halo accretion}},\ }\href
  {https://doi.org/10.1103/PhysRevResearch.2.023204} {\bibfield  {journal}
  {\bibinfo  {journal} {Phys. Rev. Res.}\ }\textbf {\bibinfo {volume} {2}},\
  \bibinfo {pages} {023204} (\bibinfo {year} {2020})},\ \Eprint
  {https://arxiv.org/abs/2002.10771} {arXiv:2002.10771 [astro-ph.CO]}
  \BibitemShut {NoStop}%
\bibitem [{\citenamefont {Ullio}\ \emph {et~al.}(2001)\citenamefont {Ullio},
  \citenamefont {Zhao},\ and\ \citenamefont {Kamionkowski}}]{Ullio:2001fb}%
  \BibitemOpen
  \bibfield  {author} {\bibinfo {author} {\bibfnamefont {P.}~\bibnamefont
  {Ullio}}, \bibinfo {author} {\bibfnamefont {H.}~\bibnamefont {Zhao}},\ and\
  \bibinfo {author} {\bibfnamefont {M.}~\bibnamefont {Kamionkowski}},\
  }\bibfield  {title} {\bibinfo {title} {{A Dark matter spike at the galactic
  center?}},\ }\href {https://doi.org/10.1103/PhysRevD.64.043504} {\bibfield
  {journal} {\bibinfo  {journal} {Phys. Rev. D}\ }\textbf {\bibinfo {volume}
  {64}},\ \bibinfo {pages} {043504} (\bibinfo {year} {2001})},\ \Eprint
  {https://arxiv.org/abs/astro-ph/0101481} {arXiv:astro-ph/0101481}
  \BibitemShut {NoStop}%
\bibitem [{\citenamefont {Shapiro}\ and\ \citenamefont
  {Heggie}(2022)}]{Shapiro:2022prq}%
  \BibitemOpen
  \bibfield  {author} {\bibinfo {author} {\bibfnamefont {S.~L.}\ \bibnamefont
  {Shapiro}}\ and\ \bibinfo {author} {\bibfnamefont {D.~C.}\ \bibnamefont
  {Heggie}},\ }\bibfield  {title} {\bibinfo {title} {{Effect of stars on the
  dark matter spike around a black hole: A tale of two treatments}},\ }\href
  {https://doi.org/10.1103/PhysRevD.106.043018} {\bibfield  {journal} {\bibinfo
   {journal} {Phys. Rev. D}\ }\textbf {\bibinfo {volume} {106}},\ \bibinfo
  {pages} {043018} (\bibinfo {year} {2022})},\ \Eprint
  {https://arxiv.org/abs/2209.08105} {arXiv:2209.08105 [astro-ph.GA]}
  \BibitemShut {NoStop}%
\bibitem [{\citenamefont {Fields}\ \emph {et~al.}(2014)\citenamefont {Fields},
  \citenamefont {Shapiro},\ and\ \citenamefont {Shelton}}]{Fields:2014pia}%
  \BibitemOpen
  \bibfield  {author} {\bibinfo {author} {\bibfnamefont {B.~D.}\ \bibnamefont
  {Fields}}, \bibinfo {author} {\bibfnamefont {S.~L.}\ \bibnamefont
  {Shapiro}},\ and\ \bibinfo {author} {\bibfnamefont {J.}~\bibnamefont
  {Shelton}},\ }\bibfield  {title} {\bibinfo {title} {{Galactic Center
  Gamma-Ray Excess from Dark Matter Annihilation: Is There A Black Hole
  Spike?}},\ }\href {https://doi.org/10.1103/PhysRevLett.113.151302} {\bibfield
   {journal} {\bibinfo  {journal} {Phys. Rev. Lett.}\ }\textbf {\bibinfo
  {volume} {113}},\ \bibinfo {pages} {151302} (\bibinfo {year} {2014})},\
  \Eprint {https://arxiv.org/abs/1406.4856} {arXiv:1406.4856 [astro-ph.HE]}
  \BibitemShut {NoStop}%
\bibitem [{\citenamefont {Shapiro}\ and\ \citenamefont
  {Shelton}(2016)}]{Shapiro:2016ypb}%
  \BibitemOpen
  \bibfield  {author} {\bibinfo {author} {\bibfnamefont {S.~L.}\ \bibnamefont
  {Shapiro}}\ and\ \bibinfo {author} {\bibfnamefont {J.}~\bibnamefont
  {Shelton}},\ }\bibfield  {title} {\bibinfo {title} {{Weak annihilation cusp
  inside the dark matter spike about a black hole}},\ }\href
  {https://doi.org/10.1103/PhysRevD.93.123510} {\bibfield  {journal} {\bibinfo
  {journal} {Phys. Rev. D}\ }\textbf {\bibinfo {volume} {93}},\ \bibinfo
  {pages} {123510} (\bibinfo {year} {2016})},\ \Eprint
  {https://arxiv.org/abs/1606.01248} {arXiv:1606.01248 [astro-ph.HE]}
  \BibitemShut {NoStop}%
\bibitem [{\citenamefont {{Rybicki}}\ and\ \citenamefont
  {{Lightman}}(1979)}]{1979rpa..book.....R}%
  \BibitemOpen
  \bibfield  {author} {\bibinfo {author} {\bibfnamefont {G.~B.}\ \bibnamefont
  {{Rybicki}}}\ and\ \bibinfo {author} {\bibfnamefont {A.~P.}\ \bibnamefont
  {{Lightman}}},\ }\href@noop {} {\emph {\bibinfo {title} {{Radiative processes
  in astrophysics}}}}\ (\bibinfo {year} {1979})\BibitemShut {NoStop}%
\bibitem [{\citenamefont {{Ghisellini}}(2013)}]{2013LNP...873.....G}%
  \BibitemOpen
  \bibfield  {author} {\bibinfo {author} {\bibfnamefont {G.}~\bibnamefont
  {{Ghisellini}}},\ }\href {https://doi.org/10.1007/978-3-319-00612-3} {\emph
  {\bibinfo {title} {{Radiative Processes in High Energy Astrophysics}}}},\
  Vol.\ \bibinfo {volume} {873}\ (\bibinfo {year} {2013})\BibitemShut {NoStop}%
\bibitem [{\citenamefont {Slatyer}\ and\ \citenamefont
  {Wu}(2017)}]{Slatyer:2016qyl}%
  \BibitemOpen
  \bibfield  {author} {\bibinfo {author} {\bibfnamefont {T.~R.}\ \bibnamefont
  {Slatyer}}\ and\ \bibinfo {author} {\bibfnamefont {C.-L.}\ \bibnamefont
  {Wu}},\ }\bibfield  {title} {\bibinfo {title} {{General Constraints on Dark
  Matter Decay from the Cosmic Microwave Background}},\ }\href
  {https://doi.org/10.1103/PhysRevD.95.023010} {\bibfield  {journal} {\bibinfo
  {journal} {Phys. Rev. D}\ }\textbf {\bibinfo {volume} {95}},\ \bibinfo
  {pages} {023010} (\bibinfo {year} {2017})},\ \Eprint
  {https://arxiv.org/abs/1610.06933} {arXiv:1610.06933 [astro-ph.CO]}
  \BibitemShut {NoStop}%
\bibitem [{\citenamefont {Slatyer}()}]{Slatyer:2017sev}%
  \BibitemOpen
  \bibfield  {author} {\bibinfo {author} {\bibfnamefont {T.~R.}\ \bibnamefont
  {Slatyer}},\ }\bibfield  {title} {\bibinfo {title} {{Indirect Detection of
  Dark Matter}},\ }\href@noop {} {\ }\Eprint {https://arxiv.org/abs/1710.05137}
  {arXiv:1710.05137 [hep-ph]} \BibitemShut {NoStop}%
\bibitem [{\citenamefont {Liu}\ \emph {et~al.}(2021)\citenamefont {Liu},
  \citenamefont {Qin}, \citenamefont {Ridgway},\ and\ \citenamefont
  {Slatyer}}]{Liu:2020wqz}%
  \BibitemOpen
  \bibfield  {author} {\bibinfo {author} {\bibfnamefont {H.}~\bibnamefont
  {Liu}}, \bibinfo {author} {\bibfnamefont {W.}~\bibnamefont {Qin}}, \bibinfo
  {author} {\bibfnamefont {G.~W.}\ \bibnamefont {Ridgway}},\ and\ \bibinfo
  {author} {\bibfnamefont {T.~R.}\ \bibnamefont {Slatyer}},\ }\bibfield
  {title} {\bibinfo {title} {{Lyman-\ensuremath{\alpha} constraints on cosmic
  heating from dark matter annihilation and decay}},\ }\href
  {https://doi.org/10.1103/PhysRevD.104.043514} {\bibfield  {journal} {\bibinfo
   {journal} {Phys. Rev. D}\ }\textbf {\bibinfo {volume} {104}},\ \bibinfo
  {pages} {043514} (\bibinfo {year} {2021})},\ \Eprint
  {https://arxiv.org/abs/2008.01084} {arXiv:2008.01084 [astro-ph.CO]}
  \BibitemShut {NoStop}%
\bibitem [{\citenamefont {Boudaud}\ \emph {et~al.}(2017)\citenamefont
  {Boudaud}, \citenamefont {Lavalle},\ and\ \citenamefont
  {Salati}}]{Boudaud:2016mos}%
  \BibitemOpen
  \bibfield  {author} {\bibinfo {author} {\bibfnamefont {M.}~\bibnamefont
  {Boudaud}}, \bibinfo {author} {\bibfnamefont {J.}~\bibnamefont {Lavalle}},\
  and\ \bibinfo {author} {\bibfnamefont {P.}~\bibnamefont {Salati}},\
  }\bibfield  {title} {\bibinfo {title} {{Novel cosmic-ray electron and
  positron constraints on MeV dark matter particles}},\ }\href
  {https://doi.org/10.1103/PhysRevLett.119.021103} {\bibfield  {journal}
  {\bibinfo  {journal} {Phys. Rev. Lett.}\ }\textbf {\bibinfo {volume} {119}},\
  \bibinfo {pages} {021103} (\bibinfo {year} {2017})},\ \Eprint
  {https://arxiv.org/abs/1612.07698} {arXiv:1612.07698 [astro-ph.HE]}
  \BibitemShut {NoStop}%
\bibitem [{\citenamefont {Boudaud}(2021)}]{Boudaud:2021zzd}%
  \BibitemOpen
  \bibfield  {author} {\bibinfo {author} {\bibfnamefont {M.}~\bibnamefont
  {Boudaud}},\ }\bibfield  {title} {\bibinfo {title} {{Voyager probing Dark
  Matter}},\ }\href {https://doi.org/10.22323/1.358.0512} {\bibfield  {journal}
  {\bibinfo  {journal} {PoS}\ }\textbf {\bibinfo {volume} {ICRC2019}},\
  \bibinfo {pages} {512} (\bibinfo {year} {2021})}\BibitemShut {NoStop}%
\bibitem [{\citenamefont {Boudaud}\ and\ \citenamefont
  {Cirelli}(2019)}]{Boudaud:2018hqb}%
  \BibitemOpen
  \bibfield  {author} {\bibinfo {author} {\bibfnamefont {M.}~\bibnamefont
  {Boudaud}}\ and\ \bibinfo {author} {\bibfnamefont {M.}~\bibnamefont
  {Cirelli}},\ }\bibfield  {title} {\bibinfo {title} {{Voyager 1 $e^\pm$
  Further Constrain Primordial Black Holes as Dark Matter}},\ }\href
  {https://doi.org/10.1103/PhysRevLett.122.041104} {\bibfield  {journal}
  {\bibinfo  {journal} {Phys. Rev. Lett.}\ }\textbf {\bibinfo {volume} {122}},\
  \bibinfo {pages} {041104} (\bibinfo {year} {2019})},\ \Eprint
  {https://arxiv.org/abs/1807.03075} {arXiv:1807.03075 [astro-ph.HE]}
  \BibitemShut {NoStop}%
\bibitem [{\citenamefont {Enoto}\ \emph {et~al.}(2019)\citenamefont {Enoto},
  \citenamefont {Kisaka},\ and\ \citenamefont {Shibata}}]{Enoto:2019vcg}%
  \BibitemOpen
  \bibfield  {author} {\bibinfo {author} {\bibfnamefont {T.}~\bibnamefont
  {Enoto}}, \bibinfo {author} {\bibfnamefont {S.}~\bibnamefont {Kisaka}},\ and\
  \bibinfo {author} {\bibfnamefont {S.}~\bibnamefont {Shibata}},\ }\bibfield
  {title} {\bibinfo {title} {{Observational diversity of magnetized neutron
  stars}},\ }\href {https://doi.org/10.1088/1361-6633/ab3def} {\bibfield
  {journal} {\bibinfo  {journal} {Rept. Prog. Phys.}\ }\textbf {\bibinfo
  {volume} {82}},\ \bibinfo {pages} {106901} (\bibinfo {year}
  {2019})}\BibitemShut {NoStop}%
\bibitem [{\citenamefont {Cohen}\ \emph {et~al.}(2017)\citenamefont {Cohen},
  \citenamefont {Murase}, \citenamefont {Rodd}, \citenamefont {Safdi},\ and\
  \citenamefont {Soreq}}]{Cohen:2016uyg}%
  \BibitemOpen
  \bibfield  {author} {\bibinfo {author} {\bibfnamefont {T.}~\bibnamefont
  {Cohen}}, \bibinfo {author} {\bibfnamefont {K.}~\bibnamefont {Murase}},
  \bibinfo {author} {\bibfnamefont {N.~L.}\ \bibnamefont {Rodd}}, \bibinfo
  {author} {\bibfnamefont {B.~R.}\ \bibnamefont {Safdi}},\ and\ \bibinfo
  {author} {\bibfnamefont {Y.}~\bibnamefont {Soreq}},\ }\bibfield  {title}
  {\bibinfo {title} {{\ensuremath{\gamma} -ray Constraints on Decaying Dark
  Matter and Implications for IceCube}},\ }\href
  {https://doi.org/10.1103/PhysRevLett.119.021102} {\bibfield  {journal}
  {\bibinfo  {journal} {Phys. Rev. Lett.}\ }\textbf {\bibinfo {volume} {119}},\
  \bibinfo {pages} {021102} (\bibinfo {year} {2017})},\ \Eprint
  {https://arxiv.org/abs/1612.05638} {arXiv:1612.05638 [hep-ph]} \BibitemShut
  {NoStop}%
\bibitem [{\citenamefont {{Koglin}}\ \emph {et~al.}(2005)\citenamefont
  {{Koglin}}, \citenamefont {{Christensen}}, \citenamefont {{Craig}},
  \citenamefont {{Decker}}, \citenamefont {{Hailey}}, \citenamefont
  {{Harrison}}, \citenamefont {{Hawthorn}}, \citenamefont {{Jensen}},
  \citenamefont {{Madsen}}, \citenamefont {{Stern}}, \citenamefont {{Tajiri}},\
  and\ \citenamefont {{Taylor}}}]{2005SPIE.5900..266K}%
  \BibitemOpen
  \bibfield  {author} {\bibinfo {author} {\bibfnamefont {J.~E.}\ \bibnamefont
  {{Koglin}}}, \bibinfo {author} {\bibfnamefont {F.~E.}\ \bibnamefont
  {{Christensen}}}, \bibinfo {author} {\bibfnamefont {W.~W.}\ \bibnamefont
  {{Craig}}}, \bibinfo {author} {\bibfnamefont {T.~R.}\ \bibnamefont
  {{Decker}}}, \bibinfo {author} {\bibfnamefont {C.~J.}\ \bibnamefont
  {{Hailey}}}, \bibinfo {author} {\bibfnamefont {F.~A.}\ \bibnamefont
  {{Harrison}}}, \bibinfo {author} {\bibfnamefont {C.}~\bibnamefont
  {{Hawthorn}}}, \bibinfo {author} {\bibfnamefont {C.~P.}\ \bibnamefont
  {{Jensen}}}, \bibinfo {author} {\bibfnamefont {K.~K.}\ \bibnamefont
  {{Madsen}}}, \bibinfo {author} {\bibfnamefont {M.}~\bibnamefont {{Stern}}},
  \bibinfo {author} {\bibfnamefont {G.}~\bibnamefont {{Tajiri}}},\ and\
  \bibinfo {author} {\bibfnamefont {M.~D.}\ \bibnamefont {{Taylor}}},\
  }\bibfield  {title} {\bibinfo {title} {{NuSTAR hard x-ray optics}},\ }in\
  \href {https://doi.org/10.1117/12.618601} {\emph {\bibinfo {booktitle}
  {Optics for EUV, X-Ray, and Gamma-Ray Astronomy II}}},\ \bibinfo {series}
  {Society of Photo-Optical Instrumentation Engineers (SPIE) Conference
  Series}, Vol.\ \bibinfo {volume} {5900},\ \bibinfo {editor} {edited by\
  \bibinfo {editor} {\bibfnamefont {O.}~\bibnamefont {{Citterio}}}\ and\
  \bibinfo {editor} {\bibfnamefont {S.~L.}\ \bibnamefont {{O'Dell}}}}\
  (\bibinfo {year} {2005})\ pp.\ \bibinfo {pages} {266--275}\BibitemShut
  {NoStop}%
\bibitem [{\citenamefont {{Ferrando}}(2002)}]{2002sf2a.conf..271F}%
  \BibitemOpen
  \bibfield  {author} {\bibinfo {author} {\bibfnamefont {P.}~\bibnamefont
  {{Ferrando}}},\ }\bibfield  {title} {\bibinfo {title} {{SIMBOL-X, an X-ray
  telescope for the 0.5-70 keV range}},\ }in\ \href
  {https://doi.org/10.48550/arXiv.astro-ph/0210229} {\emph {\bibinfo
  {booktitle} {SF2A-2002: Semaine de l'Astrophysique Francaise}}},\ \bibinfo
  {editor} {edited by\ \bibinfo {editor} {\bibfnamefont {F.}~\bibnamefont
  {{Combes}}}\ and\ \bibinfo {editor} {\bibfnamefont {D.}~\bibnamefont
  {{Barret}}}}\ (\bibinfo {year} {2002})\ p.\ \bibinfo {pages} {271},\ \Eprint
  {https://arxiv.org/abs/astro-ph/0210229} {arXiv:astro-ph/0210229 [astro-ph]}
  \BibitemShut {NoStop}%
\bibitem [{\citenamefont {Lucchetta}\ \emph {et~al.}(2022)\citenamefont
  {Lucchetta}, \citenamefont {Ackermann}, \citenamefont {Berge},\ and\
  \citenamefont {B\"uhler}}]{Lucchetta:2022nrm}%
  \BibitemOpen
  \bibfield  {author} {\bibinfo {author} {\bibfnamefont {G.}~\bibnamefont
  {Lucchetta}}, \bibinfo {author} {\bibfnamefont {M.}~\bibnamefont
  {Ackermann}}, \bibinfo {author} {\bibfnamefont {D.}~\bibnamefont {Berge}},\
  and\ \bibinfo {author} {\bibfnamefont {R.}~\bibnamefont {B\"uhler}},\
  }\bibfield  {title} {\bibinfo {title} {{Introducing the MeVCube concept: a
  CubeSat for MeV observations}},\ }\href
  {https://doi.org/10.1088/1475-7516/2022/08/013} {\bibfield  {journal}
  {\bibinfo  {journal} {JCAP}\ }\textbf {\bibinfo {volume} {08}}\bibfield
  {number} {\bibinfo  {number} { (08)},\ \bibinfo {pages} {013}},\ }\Eprint
  {https://arxiv.org/abs/2204.01325} {arXiv:2204.01325 [astro-ph.IM]}
  \BibitemShut {NoStop}%
\bibitem [{\citenamefont {Yamamoto}\ \emph {et~al.}(2023)\citenamefont
  {Yamamoto}, \citenamefont {Inui}, \citenamefont {Tada},\ and\ \citenamefont
  {Yokoyama}}]{Yamamoto:2023tsr}%
  \BibitemOpen
  \bibfield  {author} {\bibinfo {author} {\bibfnamefont {T.~S.}\ \bibnamefont
  {Yamamoto}}, \bibinfo {author} {\bibfnamefont {R.}~\bibnamefont {Inui}},
  \bibinfo {author} {\bibfnamefont {Y.}~\bibnamefont {Tada}},\ and\ \bibinfo
  {author} {\bibfnamefont {S.}~\bibnamefont {Yokoyama}},\ }\bibfield  {title}
  {\bibinfo {title} {{Prospects of detection of subsolar mass primordial black
  hole and white dwarf binary mergers}},\ }\href@noop {} {\  (\bibinfo {year}
  {2023})},\ \Eprint {https://arxiv.org/abs/2401.00044} {arXiv:2401.00044
  [gr-qc]} \BibitemShut {NoStop}%
\bibitem [{\citenamefont {Nelemans}\ \emph {et~al.}(2001)\citenamefont
  {Nelemans}, \citenamefont {Yungelson},\ and\ \citenamefont
  {Portegies~Zwart}}]{Nelemans:2001hp}%
  \BibitemOpen
  \bibfield  {author} {\bibinfo {author} {\bibfnamefont {G.}~\bibnamefont
  {Nelemans}}, \bibinfo {author} {\bibfnamefont {L.~R.}\ \bibnamefont
  {Yungelson}},\ and\ \bibinfo {author} {\bibfnamefont {S.~F.}\ \bibnamefont
  {Portegies~Zwart}},\ }\bibfield  {title} {\bibinfo {title} {{The
  gravitational wave signal from the galactic disk population of binaries
  containing two compact objects}},\ }\href
  {https://doi.org/10.1051/0004-6361:20010683} {\bibfield  {journal} {\bibinfo
  {journal} {Astron. Astrophys.}\ }\textbf {\bibinfo {volume} {375}},\ \bibinfo
  {pages} {890} (\bibinfo {year} {2001})},\ \Eprint
  {https://arxiv.org/abs/astro-ph/0105221} {arXiv:astro-ph/0105221}
  \BibitemShut {NoStop}%
\bibitem [{\citenamefont {Maguire}\ \emph {et~al.}(2020)\citenamefont
  {Maguire}, \citenamefont {Eracleous}, \citenamefont {Jonker}, \citenamefont
  {MacLeod},\ and\ \citenamefont {Rosswog}}]{Maguire:2020lad}%
  \BibitemOpen
  \bibfield  {author} {\bibinfo {author} {\bibfnamefont {K.}~\bibnamefont
  {Maguire}}, \bibinfo {author} {\bibfnamefont {M.}~\bibnamefont {Eracleous}},
  \bibinfo {author} {\bibfnamefont {P.~G.}\ \bibnamefont {Jonker}}, \bibinfo
  {author} {\bibfnamefont {M.}~\bibnamefont {MacLeod}},\ and\ \bibinfo {author}
  {\bibfnamefont {S.}~\bibnamefont {Rosswog}},\ }\bibfield  {title} {\bibinfo
  {title} {{Tidal Disruptions of White Dwarfs: Theoretical Models and
  Observational Prospects}},\ }\href
  {https://doi.org/10.1007/s11214-020-00661-2} {\bibfield  {journal} {\bibinfo
  {journal} {Space Sci. Rev.}\ }\textbf {\bibinfo {volume} {216}},\ \bibinfo
  {pages} {39} (\bibinfo {year} {2020})},\ \Eprint
  {https://arxiv.org/abs/2004.00146} {arXiv:2004.00146 [astro-ph.HE]}
  \BibitemShut {NoStop}%
\bibitem [{\citenamefont {Wang}\ \emph {et~al.}(2021)\citenamefont {Wang},
  \citenamefont {Stephan}, \citenamefont {Naoz}, \citenamefont {Hoang},\ and\
  \citenamefont {Breivik}}]{Wang:2020jsx}%
  \BibitemOpen
  \bibfield  {author} {\bibinfo {author} {\bibfnamefont {H.}~\bibnamefont
  {Wang}}, \bibinfo {author} {\bibfnamefont {A.~P.}\ \bibnamefont {Stephan}},
  \bibinfo {author} {\bibfnamefont {S.}~\bibnamefont {Naoz}}, \bibinfo {author}
  {\bibfnamefont {B.-M.}\ \bibnamefont {Hoang}},\ and\ \bibinfo {author}
  {\bibfnamefont {K.}~\bibnamefont {Breivik}},\ }\bibfield  {title} {\bibinfo
  {title} {{Gravitational-wave Signatures from Compact Object Binaries in the
  Galactic Center}},\ }\href {https://doi.org/10.3847/1538-4357/ac088d}
  {\bibfield  {journal} {\bibinfo  {journal} {Astrophys. J.}\ }\textbf
  {\bibinfo {volume} {917}},\ \bibinfo {pages} {76} (\bibinfo {year} {2021})},\
  \Eprint {https://arxiv.org/abs/2010.15841} {arXiv:2010.15841 [astro-ph.HE]}
  \BibitemShut {NoStop}%
\bibitem [{\citenamefont {Xuan}\ \emph {et~al.}(2023)\citenamefont {Xuan},
  \citenamefont {Naoz},\ and\ \citenamefont {Chen}}]{Xuan:2022qkw}%
  \BibitemOpen
  \bibfield  {author} {\bibinfo {author} {\bibfnamefont {Z.}~\bibnamefont
  {Xuan}}, \bibinfo {author} {\bibfnamefont {S.}~\bibnamefont {Naoz}},\ and\
  \bibinfo {author} {\bibfnamefont {X.}~\bibnamefont {Chen}},\ }\bibfield
  {title} {\bibinfo {title} {{Detecting accelerating eccentric binaries in the
  LISA band}},\ }\href {https://doi.org/10.1103/PhysRevD.107.043009} {\bibfield
   {journal} {\bibinfo  {journal} {Phys. Rev. D}\ }\textbf {\bibinfo {volume}
  {107}},\ \bibinfo {pages} {043009} (\bibinfo {year} {2023})},\ \Eprint
  {https://arxiv.org/abs/2210.03129} {arXiv:2210.03129 [astro-ph.HE]}
  \BibitemShut {NoStop}%
\bibitem [{\citenamefont {{Sesana}}\ \emph {et~al.}(2008)\citenamefont
  {{Sesana}}, \citenamefont {{Vecchio}}, \citenamefont {{Eracleous}},\ and\
  \citenamefont {{Sigurdsson}}}]{2008MNRAS.391..718S}%
  \BibitemOpen
  \bibfield  {author} {\bibinfo {author} {\bibfnamefont {A.}~\bibnamefont
  {{Sesana}}}, \bibinfo {author} {\bibfnamefont {A.}~\bibnamefont {{Vecchio}}},
  \bibinfo {author} {\bibfnamefont {M.}~\bibnamefont {{Eracleous}}},\ and\
  \bibinfo {author} {\bibfnamefont {S.}~\bibnamefont {{Sigurdsson}}},\
  }\bibfield  {title} {\bibinfo {title} {{Observing white dwarfs orbiting
  massive black holes in the gravitational wave and electro-magnetic window}},\
  }\href {https://doi.org/10.1111/j.1365-2966.2008.13904.x} {\bibfield
  {journal} {\bibinfo  {journal} {\mnras}\ }\textbf {\bibinfo {volume} {391}},\
  \bibinfo {pages} {718} (\bibinfo {year} {2008})},\ \Eprint
  {https://arxiv.org/abs/0806.0624} {arXiv:0806.0624 [astro-ph]} \BibitemShut
  {NoStop}%
\bibitem [{\citenamefont {Seoane}\ \emph {et~al.}(2023)\citenamefont {Seoane}
  \emph {et~al.}}]{LISA:2022yao}%
  \BibitemOpen
  \bibfield  {author} {\bibinfo {author} {\bibfnamefont {P.~A.}\ \bibnamefont
  {Seoane}} \emph {et~al.} (\bibinfo {collaboration} {LISA}),\ }\bibfield
  {title} {\bibinfo {title} {{Astrophysics with the Laser Interferometer Space
  Antenna}},\ }\href {https://doi.org/10.1007/s41114-022-00041-y} {\bibfield
  {journal} {\bibinfo  {journal} {Living Rev. Rel.}\ }\textbf {\bibinfo
  {volume} {26}},\ \bibinfo {pages} {2} (\bibinfo {year} {2023})},\ \Eprint
  {https://arxiv.org/abs/2203.06016} {arXiv:2203.06016 [gr-qc]} \BibitemShut
  {NoStop}%
\bibitem [{\citenamefont {Shao}\ and\ \citenamefont {Li}(2021)}]{Shao:2021dbg}%
  \BibitemOpen
  \bibfield  {author} {\bibinfo {author} {\bibfnamefont {Y.}~\bibnamefont
  {Shao}}\ and\ \bibinfo {author} {\bibfnamefont {X.-D.}\ \bibnamefont {Li}},\
  }\bibfield  {title} {\bibinfo {title} {{Population Synthesis of Black Hole
  Binaries with Compact Star Companions}},\ }\href
  {https://doi.org/10.3847/1538-4357/ac173e} {\bibfield  {journal} {\bibinfo
  {journal} {Astrophys. J.}\ }\textbf {\bibinfo {volume} {920}},\ \bibinfo
  {pages} {81} (\bibinfo {year} {2021})},\ \Eprint
  {https://arxiv.org/abs/2107.03565} {arXiv:2107.03565 [astro-ph.HE]}
  \BibitemShut {NoStop}%
\bibitem [{\citenamefont {Qin}\ \emph {et~al.}(2023)\citenamefont {Qin},
  \citenamefont {Jiang},\ and\ \citenamefont {Chen}}]{Qin:2023tkb}%
  \BibitemOpen
  \bibfield  {author} {\bibinfo {author} {\bibfnamefont {K.}~\bibnamefont
  {Qin}}, \bibinfo {author} {\bibfnamefont {L.}~\bibnamefont {Jiang}},\ and\
  \bibinfo {author} {\bibfnamefont {W.-C.}\ \bibnamefont {Chen}},\ }\bibfield
  {title} {\bibinfo {title} {{Black Hole Ultracompact X-Ray Binaries: Galactic
  Low-frequency Gravitational Wave Sources}},\ }\href
  {https://doi.org/10.3847/1538-4357/acb340} {\bibfield  {journal} {\bibinfo
  {journal} {Astrophys. J.}\ }\textbf {\bibinfo {volume} {944}},\ \bibinfo
  {pages} {83} (\bibinfo {year} {2023})},\ \Eprint
  {https://arxiv.org/abs/2301.06243} {arXiv:2301.06243 [astro-ph.HE]}
  \BibitemShut {NoStop}%
\bibitem [{\citenamefont {{Rosswog}}\ \emph {et~al.}(2009)\citenamefont
  {{Rosswog}}, \citenamefont {{Ramirez-Ruiz}},\ and\ \citenamefont
  {{Hix}}}]{2009ApJ...695..404R}%
  \BibitemOpen
  \bibfield  {author} {\bibinfo {author} {\bibfnamefont {S.}~\bibnamefont
  {{Rosswog}}}, \bibinfo {author} {\bibfnamefont {E.}~\bibnamefont
  {{Ramirez-Ruiz}}},\ and\ \bibinfo {author} {\bibfnamefont {W.~R.}\
  \bibnamefont {{Hix}}},\ }\bibfield  {title} {\bibinfo {title} {{Tidal
  Disruption and Ignition of White Dwarfs by Moderately Massive Black Holes}},\
  }\href@noop {} {\bibfield  {journal} {\bibinfo  {journal} {\apj}\ } (\bibinfo
  {year} {2009})}\BibitemShut {NoStop}%
\bibitem [{\citenamefont {{Ye}}\ \emph {et~al.}(2023)\citenamefont {{Ye}},
  \citenamefont {{Chen}}, \citenamefont {{Zhang}}, \citenamefont {{Fan}},\ and\
  \citenamefont {{Hu}}}]{2023MNRAS.tmp.3162Y}%
  \BibitemOpen
  \bibfield  {author} {\bibinfo {author} {\bibfnamefont {C.-Q.}\ \bibnamefont
  {{Ye}}}, \bibinfo {author} {\bibfnamefont {J.-H.}\ \bibnamefont {{Chen}}},
  \bibinfo {author} {\bibfnamefont {J.-d.}\ \bibnamefont {{Zhang}}}, \bibinfo
  {author} {\bibfnamefont {H.-M.}\ \bibnamefont {{Fan}}},\ and\ \bibinfo
  {author} {\bibfnamefont {Y.-M.}\ \bibnamefont {{Hu}}},\ }\bibfield  {title}
  {\bibinfo {title} {{Observing white dwarf tidal stripping with TianQin
  gravitational wave observatory}},\ }\bibfield  {journal} {\bibinfo  {journal}
  {\mnras}\ }\href {https://doi.org/10.1093/mnras/stad3296}
  {10.1093/mnras/stad3296} (\bibinfo {year} {2023}),\ \Eprint
  {https://arxiv.org/abs/2307.08231} {arXiv:2307.08231 [astro-ph.HE]}
  \BibitemShut {NoStop}%
\bibitem [{\citenamefont {Ye}\ \emph {et~al.}(2023)\citenamefont {Ye},
  \citenamefont {Fragione},\ and\ \citenamefont {Perna}}]{Ye:2023fpb}%
  \BibitemOpen
  \bibfield  {author} {\bibinfo {author} {\bibfnamefont {C.~S.}\ \bibnamefont
  {Ye}}, \bibinfo {author} {\bibfnamefont {G.}~\bibnamefont {Fragione}},\ and\
  \bibinfo {author} {\bibfnamefont {R.}~\bibnamefont {Perna}},\ }\bibfield
  {title} {\bibinfo {title} {{On the Tidal Capture of White Dwarfs by
  Intermediate-mass Black Holes in Dense Stellar Environments}},\ }\href
  {https://doi.org/10.3847/1538-4357/ace1eb} {\bibfield  {journal} {\bibinfo
  {journal} {Astrophys. J.}\ }\textbf {\bibinfo {volume} {953}},\ \bibinfo
  {pages} {141} (\bibinfo {year} {2023})},\ \Eprint
  {https://arxiv.org/abs/2303.07375} {arXiv:2303.07375 [astro-ph.HE]}
  \BibitemShut {NoStop}%
\end{thebibliography}%
\end{document}